# JWST Observations of Starbursts: Dust Processing in the M82 Superwind

Serena A. Cronin[1], Alberto D. Bolatto[1,2], Helena M. Richie[3], Grant P. Donnelly[4], Rebecca C. Levy[5], Karl D. Gordon[5], Elizabeth Tarantino[5], Martha L. Boyer[5], Lee Armus[6], Patricia A. Arens[7], Leindert A. Boogaard[8], Daniel A. Dale[9], Keaton Donaghue[1], Bruce T. Draine[10], Sara E. Duval[4], Kimberly Emig[11], Deanne B. Fisher[7,12], Simon C. O. Glover[13], Brandon S. Hensley[14], Rodrigo Herrera-Camus[15,16], Ralf S. Klessen[13,17], Thomas S.-Y. Lai[6], Laura Lenkić[6], Adam K. Leroy[18,19], Ashley E. Lieber[20], Ilse De Looze[21], Sebastian Lopez[18,19], David S. Meier[22], Elisabeth A.C. Mills[20], Karin M. Sandstrom[23], Evan Schneider[3], Kaitlyn E. Sheriff[20], Utsav Siwakoti[20], Evan D. Skillman[24], J.D.T. Smith[4], Yu-Hsuan Teng[1], Todd A. Thompson[18,19,25], Alexander G.G.M. Tielens[1], Sylvain Veilleux[1,2], Vicente Villanueva[26,27], Fabian Walter[28], and Paul P. van der Werf[29]

[1] Department of Astronomy, University of Maryland, College Park, MD 20742, USA; cronin@umd.edu
[2] Joint Space-Science Institute, University of Maryland, College Park, MD 20742, USA
[3] Department of Physics and Astronomy, University of Pittsburgh, 3941 O'Hara Street, Pittsburgh, PA 15260, USA
[4] Department of Physics & Astronomy and Ritter Astrophysical Research Center, University of Toledo, Toledo, OH 43606, USA
[5] Space Telescope Science Institute, 3700 San Martin Drive, Baltimore, MD 21218, USA
[6] IPAC, California Institute of Technology, 1200 East California Boulevard, Pasadena, CA 91125, USA
[7] Centre for Astrophysics and Supercomputing, Swinburne University of Technology, Hawthorn, VIC 3122, Australia
[8] Leiden Observatory, Leiden University, PO Box 9513, NL-2300 RA Leiden, The Netherlands
[9] Department of Physics and Astronomy, University of Wyoming, Laramie, WY 82071, USA
[10] Princeton University, Princeton, NJ 08540, USA
[11] National Radio Astronomy Observatory, 520 Edgemont Road, Charlottesville, VA 22903, USA
[12] ARC Centre of Excellence for All Sky Astrophysics in 3 Dimensions (ASTRO 3D), Australia
[13] Universität Heidelberg, Zentrum für Astronomie, Institut für Theoretische Astrophysik, Albert-Ueberle-Str. 2, D-69120 Heidelberg, Germany
[14] Jet Propulsion Laboratory, California Institute of Technology, 4800 Oak Grove Drive, Pasadena, CA 91109, USA
[15] Departamento de Astronomía, Universidad de Concepción, Barrio Universitario, Concepción, Chile
[16] Millennium Nucleus for Galaxies (MINGAL), Concepción, Chile
[17] Universität Heidelberg, Interdisziplinäres Zentrum für Wissenschaftliches Rechnen, Im Neuenheimer Feld 205, D-69120 Heidelberg, Germany
[18] Department of Astronomy, The Ohio State University, 140 W. 18th Avenue, Columbus, OH 43210, USA
[19] Center for Cosmology and AstroParticle Physics, The Ohio State University, 191 W. Woodruff Avenue, Columbus, OH 43210, USA
[20] Department of Physics and Astronomy, University of Kansas, 1251 Wescoe Hall Drive, Lawrence, KS 66045, USA
[21] Sterrenkundig Observatorium, Ghent University, Krijgslaan 281 - S9, B-9000 Gent, Belgium
[22] New Mexico Institute of Mining and Technology, 801 Leroy Place, Socorro, NM 87801, USA
[23] Department of Astronomy & Astrophysics, University of California, San Diego, La Jolla, CA 92093, USA
[24] Minnesota Institute for Astrophysics, University of Minnesota, 116 Church Street SE, Minneapolis, MN 55455, USA
[25] Department of Physics, The Ohio State University, 191 W. Woodruff Avenue, Columbus, OH 43210, USA
[26] Instituto de Estudios Astrofísicos, Facultad de Ingeniería y Ciencias, Universidad Diego Portales, Av. Ejército Libertador 441, 8370191 Santiago, Chile
[27] Millennium Nucleus for Galaxies, MINGAL, Chile
[28] Max Planck Institut für Astronomie, Königstuhl 17, D-69117 Heidelberg, Germany
[29] Leiden Observatory, Leiden University, P.O. Box 9513, 2300 RA Leiden, The Netherlands

## Abstract

We present JWST MIRI and NIRCam imaging of the inner ∼5 kpc of the M82 superwind at $\sim0\farcs05 - 0\farcs375$ (∼0.9–6.5 pc) resolution. Targeted filters probe emission from polycyclic aromatic hydrocarbons (PAHs; F335M, F360M, F770W, F1130W) and continuum (F250M, F360M) The images reveal a network of cool wind filaments traced by PAHs. PAH surface brightness declines with the inverse square of distance to the midplane, suggesting that the incident radiation field from the starburst drives the observed PAH intensity out to ±2.5 kpc. The 3.3/11.3 and 3.3/7.7 $\mu$m band ratios show uniformity with distance from the starburst, though comparisons with mid-IR dust emission models indicate a modest shift toward larger PAHs. Outside the disk, 11.3/7.7 $\mu$m increases moderately, reflecting that PAHs become more neutral with distance from the starburst as they are exposed to a declining radiation field and ionization parameter. Overall, PAHs in the wind are consistent with standard-to-large sizes and standard-to-high ionization states. Including Spitzer and Herschel data, PAH abundance ($q_{\rm PAH}$) is set at ∼1% in the starburst and remains unchanging out to ±5 kpc off the disk. This flat $q_{\rm PAH}$ profile suggests that PAHs are shielded from the hot wind, perhaps residing in the surface layers of cool clouds, with possible replenishment from cloud interiors and enrichment of the halo from previous bursts. In this picture, clouds are not dense enough to promote PAH growth, and they likely undergo radiative cooling and mixing with the hot phase to survive the gauntlet for at least ∼20 Myr.

## 1. Introduction

The baryon cycle in galaxies is a strong regulator of star formation, a key source of gas and metals enrichment in the circumgalactic medium (CGM), and therefore critical to galaxy evolution. Galactic superwinds (or outflows) participate in the baryon cycle by launching a significant portion of a galaxy's interstellar medium (ISM) to large distances beyond the disk. The typical model for winds driven by central starbursts is that of a cone composed of a $T \sim 10^7$ K X-ray core wrapped in a $T \sim 10^4$ K H$\alpha$-emitting layer that separates the X-ray plasma from an outer shell of entrained cooler gas. Depending on the wind's energetics, this cooler material will either escape into the surrounding halo or accrete back onto the disk in a galactic fountain (T. M. Heckman et al. 1990; F. Fraternali 2017; S. Veilleux et al. 2020; T. A. Thompson & T. M. Heckman 2024, and references therein).

Cool neutral and molecular gas dominate the mass budget of galactic winds (e.g., D. S. N. Rupke & S. Veilleux 2013; A. K. Leroy et al. 2015; R. Herrera-Camus et al. 2020; D. Salak et al. 2020; B. Mazzilli Ciraulo et al. 2025), and observations of nearby systems have revealed that this material can exist out to the CGM (C. W. Engelbracht et al. 2006; P. Beirão et al. 2015; B. Mazzilli Ciraulo et al. 2025; S. Veilleux et al. 2025). Yet it remains uncertain how cool ($T \lesssim 10^4$ K) material can survive acceleration by a hot wind. In simulations, the hot wind often ablates and destroys cooler clouds on timescales shorter than the time it would take to reach the CGM (e.g., E. E. Schneider & B. E. Robertson 2017; M. Gronke & S. P. Oh 2018; T.-E. Rathjen et al. 2021; M. W. Abruzzo et al. 2022; R. J. Farber & M. Gronke 2022; Z. Chen & S. P. Oh 2024; H. M. Richie et al. 2024). Invoking radiative cooling and turbulent mixing between the hot and cool phases may be a solution to this cool-cloud acceleration problem (M. Gronke & S. P. Oh 2018; D. B. Fielding & G. L. Bryan 2022). However, more complex physics such as cosmic rays (L. Armillotta et al. 2024; B. Sike et al. 2025) and magnetic fields (M. Sparre et al. 2020) cannot be ruled out as other mechanisms of wind-launching, the former possibly playing a role in the observed morphologies of clouds in, e.g., the wind of M82 (S. Lopez et al. 2025). Given these uncertainties, observations of the cool phase of galactic winds are crucial for constraining how this material may be delivered to the CGM, thereby providing insight into theoretical frameworks for cool-cloud survival.

In the local Universe, JWST has enabled polycyclic aromatic hydrocarbons (PAHs) as a probe of the cool ISM at previously unattainable resolution and sensitivity. The smallest form of interstellar dust, PAHs make up to ∼10%–20% of the total IR luminosity in local galaxies (J. D. T. Smith et al. 2007). PAH molecules are planar hexagonal carbon backbones with hydrogen atoms attached at the edges. Their heat capacities are small enough that a single UV-optical photon can heat the grain up to $T \sim 1000$ K (A. Leger & J. L. Puget 1984; K. Sellgren 1984). This heating manifests as complexes of spectral emission features at 3.3, 6.2, 7.7, 8.6, 11.3, 12.7, and 17.1 $\mu$m, which arise via modes such as C-H stretching (e.g., 3.3 $\mu$m), vibrational modes of the carbon "honeycomb" (e.g., 7.7 $\mu$m), and C-H bending of hydrogen atoms (e.g., 11.3 $\mu$m; A. G. G. M. Tielens 2008; B. T. Draine 2011).

Both observations and theory agree that the PAH emission in galaxies is dependent on environmental factors such as the spectral shape and strength of the incident radiation field (B. T. Draine et al. 2021; D. Baron et al. 2025; H. M. Richie & B. S. Hensley 2025), metallicity (B. T. Draine et al. 2007; G. Aniano et al. 2020; C. M. Whitcomb et al. 2024; C. M. Whitcomb et al. 2025), gas density (X. Zhang et al. 2025), gas temperature (E. R. Micelotta et al. 2010a), and turbulent shocks (E. R. Micelotta et al. 2010b; J. Y. Seok et al. 2014). For example, observations of H II regions and locations near active galactic nuclei (AGNs) have shown that the presence of strong shocks, hard radiation fields, and ionized gas can affect the properties of PAHs (J. Chastenet et al. 2019, 2023; T. Monfredini et al. 2019; O. V. Egorov et al. 2023, 2025; T. S.-Y. Lai et al. 2023; J. Chastenet et al. 2025; T. C. Fischer et al. 2025). Theoretical calculations suggest that, in the case of galactic winds, initial wind shocks of $v \gtrsim 100$ km s$^{-1}$ should destroy PAHs (B. T. Draine & E. E. Salpeter 1979a; E. R. Micelotta et al. 2010b). If shocks are slower, then PAHs should be destroyed via sputtering by ion-grain and electron-grain collisions over just a few thousand years in the post-shock, $T \gtrsim 10^6$ K gas (B. T. Draine & E. E. Salpeter 1979b; E. R. Micelotta et al. 2010a). Only PAHs that are well protected by dense cool clouds are theorized to survive such conditions on timescales of tens of Myr (e.g., H. M. Richie et al. 2024). For these reasons, measuring PAHs in galactic winds provides an excellent window into the cool material that survives launching by hot winds. Indeed, the recent detection of PAH emission extending out to ∼35 kpc into the halo of the Makani Galaxy demonstrates that PAHs can survive the extreme conditions of winds on long (∼100 Myr) timescales if shielded (S. Veilleux et al. 2025).

M82 is the archetypal starburst galaxy with a multiphase outflow driven by feedback from supernovae and young stellar clusters ($\lesssim$10 Myr; N. M. Förster Schreiber et al. 2001; Y. D. Mayya et al. 2008; R. C. Levy et al. 2024). The bulk of the outflow mass is in the molecular phase ($\sim 3 \times 10^8\, M_\odot$; F. Walter et al. 2002; P. Beirão et al. 2015). Combined with the angular resolution and sensitivity achieved by JWST, we can study the M82 wind on the scales of individual star-forming regions ($\lesssim$7 pc resolution) due to its proximity ($d \sim 3.6$ Mpc; W. L. Freedman et al. 1994; J. J. Dalcanton et al. 2009) and approximately edge-on geometry ($i \sim 80°$; Y. D. Mayya et al. 2005). The cool phase of the wind, as traced by PAHs and dust continuum emitting at 8 $\mu$m, extends to at least ∼6 kpc beyond the galaxy's disk (Figure 1; C. W. Engelbracht et al. 2006). Measuring PAH band ratios using the IRS instrument on Spitzer, P. Beirão et al. (2015) reported variations in PAH charge and size throughout the M82 wind, which could imply some processing of the cool phase over time. This was also the first spectroscopic evidence that PAHs can survive the M82 wind. Recently, JWST has hinted at an interplay between PAHs and multiple phases of the wind, with PAH emission correlating spatially with both ionized and neutral (atomic and molecular) gas (A. D. Bolatto et al. 2024; D. B. Fisher et al. 2025; V. Villanueva et al. 2025a, 2025b; S. Lopez et al. 2026). The complete impact that warmer phases have on the dusty cool gas remains not fully understood.

The goal of this paper is to diagnose the physical state of PAHs in the M82 wind and trace the processing of the cool phase over ∼10–20 Myr. In doing so, we build upon our team's previous work, which detected PAHs at the base of the wind (A. D. Bolatto et al. 2024; T. C. Fischer et al. 2025), investigated correlations between PAHs and various gas phases farther out (V. Villanueva et al. 2025a, 2025b; S. Lopez et al. 2026), and identified the young massive star clusters in the starburst that may contribute to the launching of this material (R. C. Levy et al. 2024). Here, we present $\sim 0\farcs05 - 0\farcs375$ (∼0.9–6.5 pc) resolution JWST observations of the inner ∼5 kpc of the wind, focusing on PAH emission at 3.3, 7.7, and 11.3 $\mu$m. The structure of this work is as follows. In Section 2, we detail these observations and data reduction and present the resulting mosaics and emission measurements in Section 3. In Section 4, we use PAH band ratios as a measurement of the physical state of PAHs over vertical distances of $\sim\pm2$ kpc. Combining the JWST data with archival Spitzer and Herschel images, we measure changes in PAH abundance in Section 5. We discuss the implications of these findings in Section 6 and summarize them in Section 7.

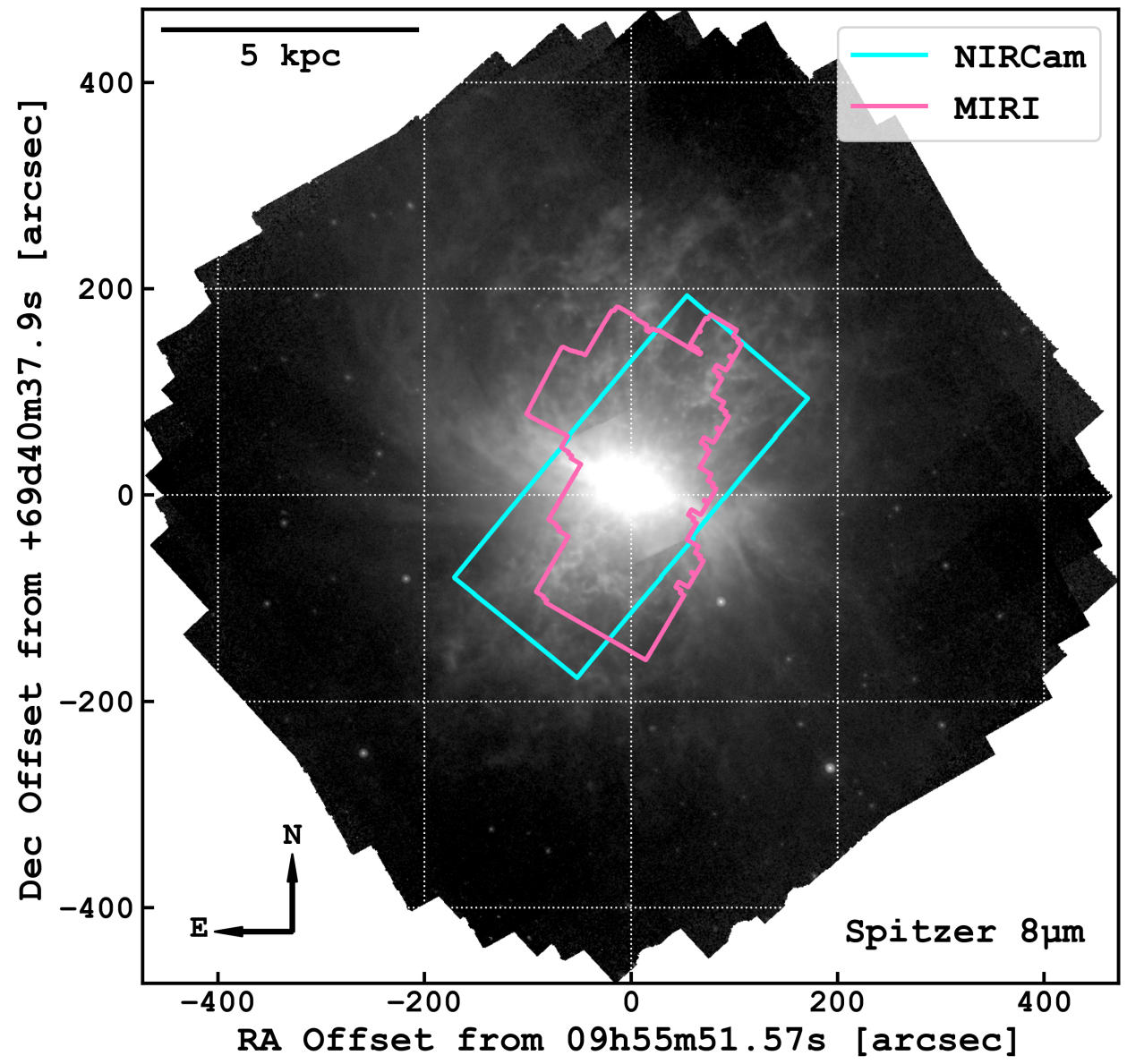

**Figure 1.** Spitzer IRAC image at 8 $\mu$m with NIRCam (blue) and MIRI (pink) footprints overlaid. The JWST observations map the inner wind along the minor axis out to ∼2.5 kpc north and south of the disk. M82 is at a distance of 3.6 Mpc (W. L. Freedman et al. 1994; J. J. Dalcanton et al. 2009) and is nearly edge-on ($i \sim 80°$; Y. D. Mayya et al. 2005).

## 2. Data

The JWST imaging presented in this work was obtained as part of Cycle 1 GO project 1701 (PI: A. Bolatto). The MIRI and NIRCam coverage areas are overlaid onto a Spitzer 8 $\mu$m image of M82 in Figure 1. These areas cover the galaxy midplane and inner ∼5 kpc of the wind.

### 2.1. NIRCam Observations

The main NIRCam observations used in this work targeted the F250M, F335M, and F360M filters. As detailed in Section 2.6, the F250M and F360M images are used to isolate the 3.3 $\mu$m PAH emission by separating it from the underlying continuum in the F335M image. Our observing program also targeted continuum and line imaging using the F140M, F164N ([Fe II]), and F212N ($H_2 \nu = 1 \rightarrow 0$) filters. While these latter three images are not used for scientific analysis in this paper, we present them in a multicolor image of the M82 wind in Figure 2.

The NIRCam observations were carried out in two parts: a `FULL` array exposure covering the wind along the galaxy minor axis, and a short `SUB640` exposure of the central starburst designed to avoid saturation. The `FULL` array setup consisted of a 1 × 1 mosaic, a `MEDIUM8` readout pattern, four `FULLBOX` primary dithers and three `SMALL-GRID-DITHER` subpixel dithers, and four groups per integration. These observations were executed on 2024 January 5 between 05:41:08.271 and 11:15:25.272 UTC, yielding a total effective exposure time of 4895.96 s (1.36 hr).

The central starburst data were obtained using a 1 × 1`SUB640` exposure with the `RAPID` readout, four `INTRAMODULEBOX` primary and four `SMALL-GRID-DITHER` subpixel dithers, and six groups per integration. These data were taken on 2022 October 18 between 00:35:36.247 and 01:38:41.533 UTC, for a total effective exposure time of 401.84 s (0.11 hr). These data were first presented in A. D. Bolatto et al. (2024) and D. B. Fisher et al. (2025).

### 2.2. MIRI Observations

This observing program targeted three filters with the MIRI imager: F560W (continuum and part of the PAH complex at 6.2 $\mu$m), F770W (continuum and the 7.7 $\mu$m PAH complex), and F1130W (continuum and the 11.3 $\mu$m PAH complex). The F770W and F1130W images were first presented in V. Villanueva et al. (2025a, 2025b) and S. Lopez et al. (2026). A composite image is shown in Figure 2.

As with NIRCam, the MIRI observations were carried out in two parts to avoid saturating the detector over the bright starburst. The wind was captured in a 1 × 4 mosaic with the `FULL` array using the `FASTR1` readout pattern, 4-point `cycling` dither, and 30 groups per integration. These observations were taken on 2023 December 31 between 15:53:20.695 and 18:47:06.720 UTC. The total effective exposure time was 2331.028 s (0.65 hr).

Using the `SUB128` subarray, the central starburst was covered with a 2 × 2 mosaic and executed with the `FASTR1` readout pattern, a `4-Point-Sets` dither pattern optimized for extended sources, and five groups per integration. These subarray observations were acquired on 2023 December 31 between 13:55:26.942 and 15:27:05.388 UTC, resulting in a total effective exposure time of 9.52 s.

The MIRI observations also required a dedicated estimate of background emission levels. We executed a 1 × 1`FULL` array exposure off-source at position (R.A., decl.) J2000 = (147°679, 69°8306). This background observation was taken using the `FASTR1` readout pattern, the `4-Point-Sets` dither optimized for extended sources, and 30 groups per integration. The total effective exposure time was 333.004 s, with the observation taken on 2024 November 16 at 06:10:47 UTC. Due to guide star acquisition failures, the background data were taken nearly a year after the science data. However, as described in Section 2.7, the calculated background levels are consistent with expectations from the JWST Backgrounds Tool.[30]

### 2.3. Pipeline Reduction of MIRI Imaging

We ran the JWST pipeline development version 1.9.5 (H. Bushouse et al. 2023) with reference files specified by CRDS context `jwst_1188.pmap` on the uncalibrated MIRI imager data products. The default parameters are assumed except for the `firstframe` step in Stage 1, where we toggle `bright_use_group1= True` for the F770W reduction only to help recover saturated sources. This disables flagging the first group in every integration as bad. There is still some saturation in the midplane due to the bright starburst, but this occurs in <1% of all pixels. For the F560W and F1130W reductions, we adopt the default parameters in all three stages of the JWST pipeline.

[30] https://github.com/spacetelescope/jwst_backgrounds

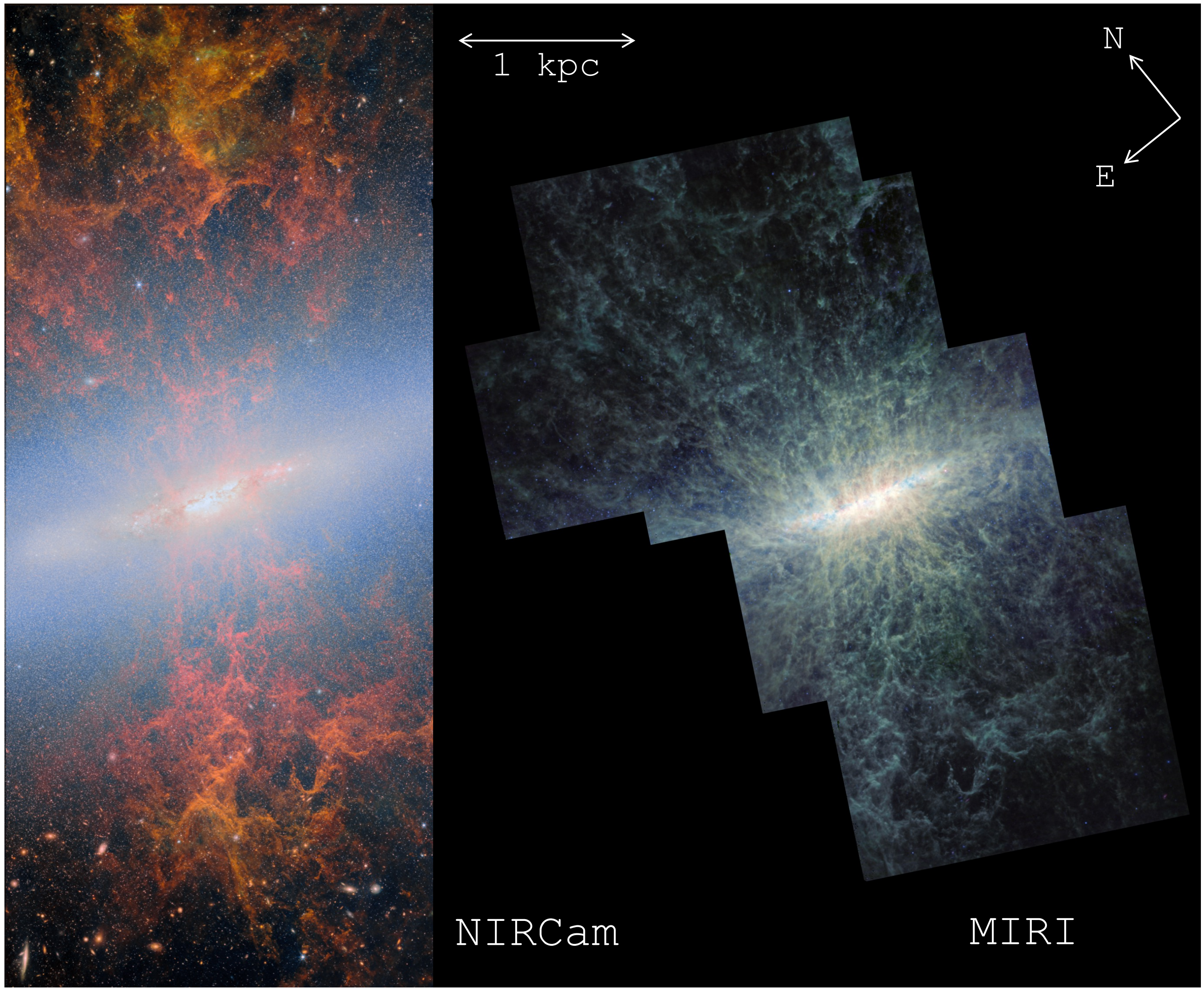

**Figure 2.** JWST imaging of the M82 superwind revealing the complex filamentary structure traced by PAH emission on ∼0.9–6.5 pc physical scales out to at least ∼2 kpc from the starburst disk. Left panel: composite image taken with the following NIRCam filters: F140M (blue; continuum), F164N (cyan; [Fe II]), F212N and F250M (green; $H_2$ and continuum), F335M (yellow; PAHs and continuum), and F360M (red; some PAHs and continuum). Right panel: composite MIRI imaging at $0\farcs375 \approx 6.5$ pc resolution (point-spread function, PSF, of F1130W). Three MIRI filters are used: F560W (red; mostly continuum with some PAHs at 6.2 $\mu$m), F770W (green; 7.7 $\mu$m PAHs and continuum), and F1130W (blue; 11.3 $\mu$m PAHs and continuum). The F770W and F1130W images contain mostly PAH emission with ∼14%–20% contribution from continuum. Credits: (left) NASA, ESA, CSA, STScI, A. Bolatto (UMD); image created by A. Pagan (STScI). (right) ESA/Webb, NASA & CSA, A. Bolatto. A composite MIRI image from ESA/Webb, NASA & CSA can be found at https://esawebb.org/images/potm2506b/.

### 2.4. Pipeline Reduction of NIRCam Imaging

We ran the JWST pipeline version 1.15.1 (H. Bushouse et al. 2024) on the uncalibrated NIRCam data products. We kept the default parameters of Stage 1 of the pipeline with two exceptions. During the ramp-fitting step, we set `suppress_one_group=False` to recover saturated sources in the first integration group. We also flagged the "snowball" features from large cosmic-ray events by setting `expand_large_events = True` in the jump step, which is now a default setting in more recent pipeline versions. Stage 2 of the pipeline was then run for all NIRCam images with the default settings. Between Stages 2 and 3, we also applied the C. J. Willott et al. (2022) algorithm to the `□_cal.fits` files to remove striping caused by $1/f$ noise that is common in NIRCam imaging. This $1/f$ noise correction works well for the main filters used in this work (F250M, F335M, and F360M), though we note that some striping persists in the narrowband filters (F164N and F212N).

Aligning the NIRCam exposures to produce a final mosaic proved tricky with a lack of bright sources and an insufficient number of stars cataloged by Gaia in the field of view. To assist the alignment, we ran Stage 3 of the JWST pipeline on the F250M `FULL` array only. The `tweakreg` parameters were heavily modified by the following: `brightest = 2000`, `SNR_threshold = 4`, `peakmax = 500`, `tolerance = 0.2`, and `expand_refcat = True`. These altered parameters resulted in a decent F250M `FULL` mosaic with minimal distortions. We then aligned this mosaic to Gaia DR3 stars using the JWST/HST Alignment Tool (`JHAT`; A. Rest et al. 2023).

To create a high-quality catalog of stars, we ran this Gaia-aligned F250M mosaic through the stellar photometry software DOLPHOT (A. E. Dolphin 2000; A. Dolphin 2016; D. R. Weisz et al. 2024). DOLPHOT also performed an alignment before generating this catalog, improving upon the previous alignments. With our catalog of stars from the F250M image, we used JHAT to produce our final NIRCam mosaics (SUB640 + FULL arrays) for each filter.

A few artifacts remain after this process. There are some mismatched backgrounds between exposures. This is most noticeable in a slanted region toward the center of the mosaic where the SUB640 exposure meets a nearby FULL exposure. There are also extraneous pixels at the edges of the images, perhaps due to cosmic rays. Modifying the jump step did not improve these regions. Finally, there are a few missing pixels also at the edges of the map that were not remedied by tweaking the outlier detection parameter. We suspect these are bad pixels in the detector. However, these few artifacts do not limit our science, and the final images show a vast improvement upon the standard JWST pipeline, resulting in sharp point sources with very minimal distortions.

### *2.5. Matching the Point-spread Function*

To properly subtract the continuum and calculate ratios of emission at different wavelengths, we must match the point-spread function (PSF) of the images. The goal is to effectively match the angular resolution across the images; in practice, because we are using circular (azimuthally averaged) PSFs, this is an approximation for JWST given its strong diffraction spikes and inherently nonaxisymmetric PSF. We follow the procedure that produced the continuum-subtracted 3.3 $\mu$m image of the central ~1 kpc of M82 in A. D. Bolatto et al. (2024). Using the 1.2.0 version of the Space Telescope PSF tool (STPSF; formerly WebbPSF), we oversample the PSFs of each image (F250M, F335M, F360M, F770W, and F1130W) to 0$''$014, and generate convolution kernels with pypher (A. Boucaud et al. 2016). A. D. Bolatto et al. (2024) found that the default regularization parameter of $10^{-4}$ works well when computing the kernels, with little to no difference in the results when experimenting with different values. This parameter stabilizes the kernels against noisy high frequencies. For MIRI, we match the F770W image to the longest wavelength image, F1130W; for NIRCam, we match F250M and F335M to F360M.

G. Aniano et al. (2011) developed a figure of merit ($W_{-}$) to check the aggressiveness of convolution kernels. Kernels that are safe to use have a figure of merit $W_{-} \leqslant 1$, while kernels with $W_{-} > 1.2$ could generate artifacts and thus should not be used. Kernels with $W_{-} \approx 1$ are considered only mildly aggressive and may be used with caution. We inspect our kernels using this figure of merit and find that all kernels are nonaggressive and safe to use, with $W_{-} = 0.08$–$0.13$.

With these kernels and the images regridded to a common pixel size (0$''$014), we convolve the MIRI images to the lowest-resolution image (F1130W; PSF = 0$''$375 = 6.5 pc) using a fast Fourier transform. We perform this convolution for the NIRCam images at the resolution of F360M (PSF = 0$''$12 = 2.1 pc). Because of the size of the resulting images, the final products used in this analysis are regridded once more to a pixel scale of 0$''$05 = 0.87 pc.

### *2.6. Continuum Subtraction of the F335M Image*

The F335M image contains both continuum emission and the 3.3 $\mu$m PAH feature. The 3.3 $\mu$m PAH feature is a complex that includes the aliphatic 3.4 $\mu$m feature and the aromatic plateau at 3.47 $\mu$m (M. Hammonds et al. 2015). Both features are weaker than the emission at 3.3 $\mu$m PAH, though the plateau may make up ~20% of the total 3.3 $\mu$m power (T. S.-Y. Lai et al. 2020). The shoulder of a broad $H_2O$ ice feature may also contaminate the 3.3 $\mu$m complex at 3.05 $\mu$m, causing absorption especially in extinguished regions (M. Hammonds et al. 2015; T. S.-Y. Lai et al. 2020; A. D. Bolatto et al. 2024).

To remove the continuum from the 3.3 $\mu$m complex, we use the prescription developed by A. D. Bolatto et al. (2024), which worked well on the SUB640 image of the base of the M82 outflow that is also included in this work. The NIRCam observations include the F250M and F360M filters in order to estimate the continuum in the F335M image. A. D. Bolatto et al. (2024) found that the F250M filter is a good measure for the continuum, free of contamination from other features. This is corroborated by spectroscopic measurements of M82 (E. Sturm et al. 2000; N. M. Förster Schreiber et al. 2001). We use the F360M filter to flank the F335M filter alongside F250M; however, F360M is likely contaminated by the longer-wavelength parts of the 3.3 $\mu$m PAH complex, including the aliphatic and aromatic features at 3.4 and 3.47 $\mu$m (M. Hammonds et al. 2015; T. S.-Y. Lai et al. 2020; A. D. Bolatto et al. 2024). These contributions to the F360M filter are correlated with PAH-bright regions as seen in F335M (K. M. Sandstrom et al. 2023). Thus, we can use an iterative process that estimates the fractional contribution of PAH emission to the F360M filter. The leftover F360M continuum is then used in tandem with the F250M image to diagnose the total underlying continuum in F335M.

We refer to A. D. Bolatto et al. (2024) for details on the full continuum-subtraction prescription, but summarize the methodology here. The iterative procedure begins with subtracting a fraction $q$ of the 3.3 $\mu$m PAH emission from the F360M image (the fraction being 0 in the first iteration). A weighted combination of the remaining 3.6 $\mu$m continuum and the emission in F250M is then used to estimate the continuum in F335M. This continuum is then subtracted from the F335M image. The process repeats, increasing the fraction of 3.3 $\mu$m PAH emission $q$ until the results converge. Our only change with this procedure is our choice of $q = 0.55$ and $R = 0.65$ (another color parameter), which we find works better than what A. D. Bolatto et al. (2024) used for solely the base of the wind. We perform this continuum subtraction at the F360M resolution. The continuum image subtracted from the F335M map can be found in Figure 11 of Appendix A. Afterwards, we convolve the 3.3 $\mu$m PAH map to match the MIRI images at the F1130W resolution (0$''$375, 6.5 pc).

### *2.7. Background and Continuum Subtraction of the MIRI Images*

Background emission estimates are essential for mid-IR observations. Because M82 is a crowded field, this leads to a lack of empty regions in the MIRI science mosaics, thus requiring a dedicated off-source background observation (see details in Section 2.2). For each image background observation, we take a large aperture ($r = 12''$) that is free of galaxies and estimate a mean background of 4.8 MJy sr$^{-1}$ and 19 MJy sr$^{-1}$ for F770W and F1130W, respectively. These values are consistent with the backgrounds estimated by S. Lopez et al. (2026). Due to guide star acquisition failures, the background and science exposures were taken approximately a year apart; however, we find that our estimate of the background is highly consistent with expectations from the JWST Backgrounds Tool. Comparison between our background estimates and this tool reveals that the zodiacal light dominates the F770W background, while a combination of zodiacal light and thermal emission dominates the F1130W background.

Our observing program was originally designed to use the F560W filter to estimate the continuum likely contributing to the F770W and F1130W filters. However, because the F560W filter also measures a variable fraction of PAH emission at 6.2 $\mu$m, we instead opt to use legacy Spitzer spectral maps to estimate the continuum. The background-subtracted IRS mapping-mode cubes for M82 were obtained from the Spitzer IRS Mapping Legacy Archive (G. P. Donnelly et al. 2025a; see also P. Beirão et al. 2008, 2015). We created a spectral map of the M82 system by mosaicking together the available cubes (program IDs: 21, 159, 30542, 50575). Using the IDL version of PAHFIT (J. D. T. Smith et al. 2007), which is specifically designed to work with IRS data, we decomposed the spectrum of each spaxel of the mosaic into separate components of emission and continuum. To properly estimate the continuum, we derived synthetic photometry from the continuum model of each pixel separately for the F770W and F1130W filters using their respective transmission curves (A. Glasse et al. 2015). We adopt the absolute JWST flux calibration from Equation (5) in K. D. Gordon et al. (2022; see also J. Koornneef et al. 1986). From the synthetic photometry, we then generated maps of the continuum component for F770W and F1130W. We note that these maps are at the IRS spatial resolution ($\sim$5″ at the long-wavelength module pixel size) and, thus, are very coarse compared to MIRI.

The resulting continuum maps for F770W and F1130W are smooth across large scales (see Figure 11 in Appendix A). We note that assuming this smooth continuum and regridding to the same pixel scale as the MIRI maps (0.″05) is an approximation, since the true continuum may vary on scales that are unresolvable by the IRS. Nonetheless, this does not introduce spurious artificial structure into the PAH maps, and so we use these continuum maps to perform a pixel-by-pixel subtraction of the F770W and F1130W images. We note that about 75% of the JWST map has corresponding coverage with the IRS. The difference in area covered is due to mismatching mosaic shapes and the IRS map covering a larger extent of the wind. For regions that are not covered by the IRS mosaic, we instead subtract the median contribution of continuum, adopting values for each morphological component: $\sim$14% (wind) and $\sim$16% (disk) for F770W, and $\sim$18% (wind) and $\sim$22% (disk) for F1130W. These calculations were previously reported in S. Lopez et al. (2026) and are in line with expectations. For reference, C. M. Whitcomb et al. (2023) used a wide set of Spitzer IRS spectra (360 from wide-area radial strips across three $z \sim 0$ galaxies, and 120 from star-forming environments in 46 SINGS galaxies) to determine that both images are dominated by PAH emission with moderate contributions from dust emission ($\sim$15% in F770W and $\sim$25% in F1130W) and small contributions from starlight ($\sim$2% in F770W and <1% in F1130W). Recent MIRI-MRS observations of edge-on galaxy NGC 891's halo (out to 1 kpc) detected upwards of $\sim$20% stellar and $\sim$10% hot dust continua in F770W, and $\sim$10% stellar and $\sim$20% hot dust continua in F1130W, and an overall a declining PAH-to-continuum ratio with distance from the disk (J. Chastenet et al. 2026).

The final 7.7 and 11.3 $\mu$m PAH emission maps show no leftover artifacts from the background and continuum subtractions (Figure 3; see Appendix A).[31]

## 3. PAH Emission

The panels of Figure 2 contain NIRCam and MIRI imaging of the inner $\sim$5 kpc of the M82 wind, respectively. These images still contain continuum and background emission. The MIRI image (right) is a composite F560W (red), F770W (green), and F1130W (blue) mosaic highlighting regions of 6.2 $\mu$m PAH + continuum, 7.7 $\mu$m PAH emission, and 11.3 $\mu$m PAH emission, respectively. The F770W and F1130W images also contain some dust and starlight continuum. Tendrils of PAH emission stretch from the galaxy midplane and to the edge of the field of view. Based on the NIRCam SUB640 imaging at the very center of the galaxy, A. D. Bolatto et al. (2024) and D. B. Fisher et al. (2025) noted that these filamentary structures originate as plumes extending radially outward from the base of the wind. At a short distance from the starburst ($\sim$300–600 pc), other structures resembling ridges and arches, which may be interpreted as bubble walls, appear to occur regularly across the images. These arcs are also notable in H$\alpha$ (S. Lopez et al. 2025).

In the left panel of Figure 2, we present a multicolor NIRCam image using the F140M (blue), F164N (cyan), F212N (green), F250M (yellow), F335M (orange), and F360M (red) filters. The continuum emission overpowers the outflow emission toward the midplane. The F335M-dominated emission traces the edges of the outflow cone across the starburst in both the north and the south. Possible dust lanes are also visible in the galaxy disk.

Figure 3 shows the background- and continuum-subtracted maps of 3.3, 7.7, and 11.3 $\mu$m emission at their native resolutions: 0.″269 $\approx$ 4.7 pc resolution for F770W and 0.″375 $\approx$ 6.5 pc resolution for F1130W. The 3.3 $\mu$m map is at the 0.″12 $\approx$ 2 pc resolution of the F360M image. Overall, we find the continuum subtraction of the 3.3 $\mu$m complex to be good, as the resulting continuum resembles a stellar disk. Some artifacts remain in the 3.3 $\mu$m map caused by an imperfect matching of the SUB640 array and the FULL array toward the center, different background levels between footprints in the north, and contributions from cosmic rays or missing pixels. The 7.7 $\mu$m map, on the other hand, only suffers from a few missing pixels at the center of the mosaic due to saturation. The aforementioned artifacts are minor and do not affect our science. The 11.3 $\mu$m map shows no visible artifacts.

The PAH surface brightness at 11.3 $\mu$m is the brightest out of the three features of our study, with only a factor of $\sim 2\times$ difference between the 11.3 and 7.7 $\mu$m features (see relative surface brightness ratios in Section 4). A similar elevation of brightness in 11.3 $\mu$m with respect to 7.7 $\mu$m has also recently been found in the halo of the edge-on spiral galaxy NGC 891 (J. Chastenet et al. 2026, submitted). Emission from the 3.3 $\mu$m complex is $\sim$10$\times$ weaker than both mid-IR features. The ratio of 3.3–11.3 $\mu$m is expected, as the 3.3 $\mu$m complex only contributes $\sim$2% of the total PAH power in star-forming galaxy disks (C. M. Whitcomb et al. 2024), including starbursts and luminous infrared galaxies (T. S.-Y. Lai et al. 2020).

[31] We note that the continuum subtraction of the NIRCam (Section 2.6) and MIRI (Section 2.7) images is performed at different physical scales, with the MIRI continuum estimated at the coarser IRS resolution and the F335M continuum estimate derived at the higher F360M resolution. To ensure that these differing physical scales do not introduce artificial trends, we tested performing the F335M continuum subtraction at the IRS resolution, and found that it does not change the PAH ratios presented throughout this work.

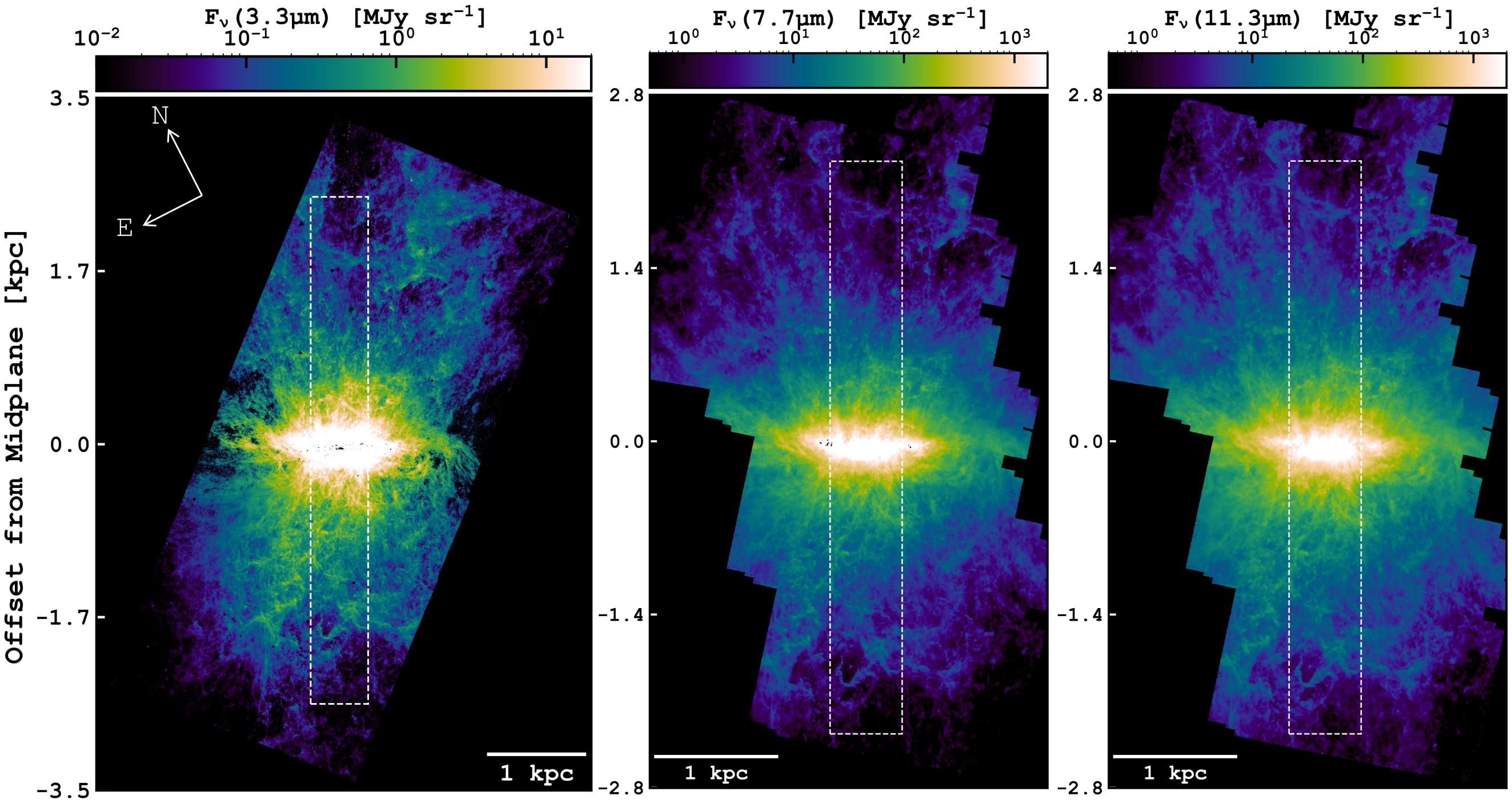

**Figure 3.** Maps of 3.3, 7.7, and 11.3 $\mu$m PAH emission in units of $F_{\nu}$ [MJy sr$^{-1}$]. These images are background- and continuum-subtracted (see Section 2), and rotated to align with the galaxy minor axis. The MIRI maps are presented at their native resolutions: 0.″269 $\approx$ 4.7 pc resolution for F770W and 0.″375 $\approx$ 6.5 pc resolution for F1130W. The 3.3 $\mu$m map is at the 0.″12 $\approx$ 2 pc resolution of the F360M image. We average emission values in a 33″-wide slice along the minor axis (white rectangles) and present them in Figure 4.

The top panel of Figure 4 shows the distribution of PAH surface brightness as a function of distance from the midplane in kiloparsecs (bottom $x$-axis), where a negative offset corresponds with south of the midplane, positive offset is toward the north, and the midplane itself is at 0 kpc. Distance from the midplane is defined as projected vertical height $z$. We convert this distance in kiloparsecs to an estimated timescale in Myr (top $x$-axis) by assuming a constant wind velocity $v$ = 200 km s$^{-1}$, which is typical for the colder phases of galactic winds (e.g., A. D. Bolatto et al. 2013) and consistent with the velocities of dust in simulations of M82-like outflows (H. M. Richie & E. E. Schneider 2026). We acknowledge that PAHs are also spatially associated with ionized gas phases (S. Lopez et al. 2026), and that the H$\alpha$-emitting phase is estimated to be moving at $\sim$500–600 km s$^{-1}$ (P. L. Shopbell & J. Bland-Hawthorn 1998). However, simulations show cold dense gas accelerates more slowly than diffuse ionized gas in galactic winds (H. M. Richie et al. 2024). Therefore, the 200 km s$^{-1}$ adopted in this work reflects the colder gas phases in which PAHs are likely embedded during their initial entrainment in the outflow. If we were to instead assume a higher cool-wind velocity, then the timescales quoted throughout this paper would decrease.

To produce the intensity profiles, we average the surface brightness over discrete rows of $z$ in a 33″-wide ($\approx$0.6 kpc) slice taken down the center of each image, aligned with the minor axis of M82 (white rectangles in Figure 3). The shaded regions represent the standard deviation to highlight the range of values being averaged. In the bottom panel of Figure 4, we divide the average $F_{\nu}$ by the inverse square function of vertical distance, $z^{-2}$, in order to remove the flux dependence on distance from the incident radiation field. This assumes stochastic heating such that mid-IR emission scales linearly with $U \propto z^{-2}$. We mask the disk between $\pm$0.1 kpc where $z^{-2}$ becomes undefined. By factoring out the dependence on distance from the incident radiation field, all three features predominantly show a uniform distribution in $F_{\nu}$ versus vertical distance. This suggests that changes in PAH surface brightness could be predominately due to the effects of distance from the incident radiation field rather than changes in the intrinsic PAH properties. We note a dip in the surface brightness profiles of all three PAH tracers between $z \approx +0.5$–1.5 kpc. This dip is due to a hole of emission in the north, which is also notable in H$\alpha$ (S. Lopez et al. 2026) and flattens out when using a larger (100″ in width) slice. Therefore, this decrease is likely an effect of the column-density of the clouds in this region.

Extinction likely plays a role in the observed PAH intensities. One of the most prominent examples is a broad silicate absorption feature at 9.7 $\mu$m (e.g., J. D. T. Smith et al. 2007), which would have its largest effect on disk emission captured by F1130W. Assuming mixed geometry and the default extinction curve in PAHFIT, we model the continuum using Spitzer spectroscopy (Section 2.7) and calculate an average absorption depth of $\tau_{9.7} \approx 1.5$ ($A_{\lambda} \sim 1.6$ mag) in the M82 disk. This value is in line with the $\tau_{9.7} = 0.3$–3.1 (median $\tau_{9.7} = 1.3$) that P. Beirão et al. (2008) calculated using the same Spitzer data, and a median $\tau_{9.7} = 1.3 \pm 0.3$ measured in the starburst with the MIRI Medium Resolution Spectroscopy Integral Field Unit (IFU) instrument, assuming screen geometry (S. Duval et al. 2026, in preparation). Beyond the silicate absorption feature, extinction exists across the entire IR spectrum. Following a B. S. Hensley & B. T. Draine (2023) extinction curve, and assuming $\tau_{9.7} \approx 1.5$ at 9.7 $\mu$m, we expect $A_\lambda \approx 1$ mag around 3.3 and 7.7 $\mu$m. We will discuss the role that extinction may play in the observed band ratios in the next section.

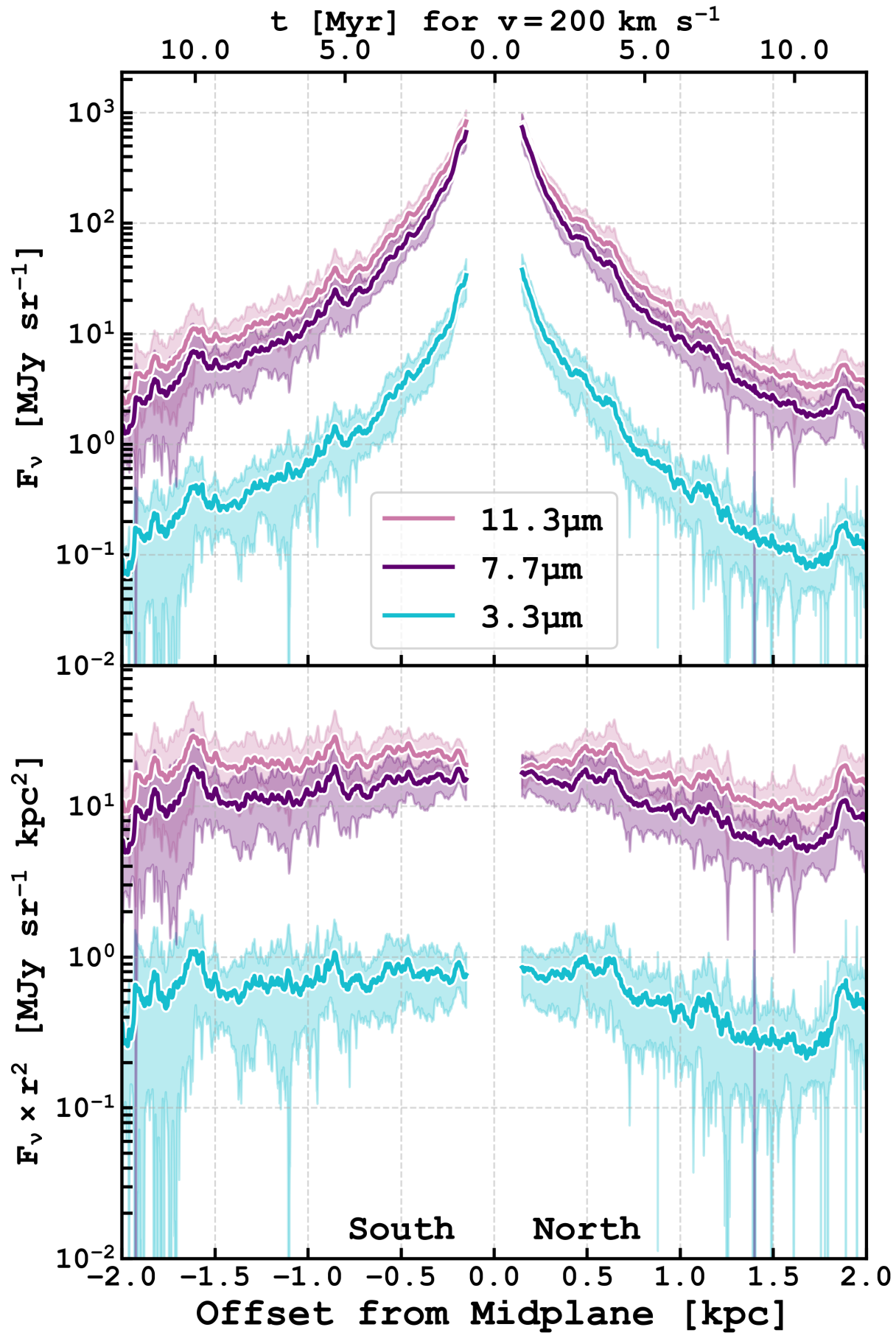

**Figure 4.** Top panel: PAH surface brightness $F_\nu$ [MJy sr$^{-1}$] as a function of projected vertical distance ($z$) from the illuminating starburst. Positive offset is the northern wind, negative offset is the southern wind, and $z \approx 0$ references the disk midplane. We measure the emission at a matched 0.″375 ≈ 6.5 pc resolution of the F1130W image. These intensity profiles are averaged over each discrete $z$ in a 33″-wide rectangular slice down the center of the images (Figure 3). The shaded regions are the standard deviation of values being averaged. The timescale on the top axis assumes material is moving at 200 km s$^{-1}$. Bottom panel: the same as above, but removing the inverse square function of vertical distance ($z^{-2}$) from the surface brightness, and masking between ±0.1 kpc where $z$ becomes undefined. The PAH surface brightness profiles are largely uniform over the inner ±2 kpc of the wind, indicating that distance from incident radiation field plays a dominant role in changes to the observed surface brightness. The drop beyond 0.5 kpc in the north is due to a hole of emission (noticeable in Figure 3), and levels out when averaging over a larger slice.

We verify that these results are independent of the size and shape of the slice by deriving profiles using a larger (100″-wide) rectangle, and a physically motivated cone with an opening angle of 60° (F. Walter et al. 2002) and its apex situated at the galactic center. Given that these results are consistent, we adopt the rectangular slice for simplicity, and prefer the less-wide rectangle to minimize emission from the disk. Additionally, we note that this analysis is done in projected vertical ($z$) geometry instead of projected radial geometry. Assuming most of the emission arises from the surface layer of a cone that lies in the plane of the sky, any mixing of multiple projected radii per $z$ bin is negligible.

## 4. PAH Ratios

The relative strength between different PAH features is a diagnostic of the state of PAHs, including size and ionization. In this section, we employ the 11.3/7.7, 3.3/7.7, and 3.3/11.3 $\mu$m band ratios to track PAH processing throughout the wind. The 11.3/7.7 $\mu$m ratio is most sensitive to PAH ionization with some dependence on size, with higher values indicating more neutral (and perhaps larger) PAHs. The 3.3/7.7 and 3.3/11.3 $\mu$m ratios, on the other hand, are mostly sensitive to the size distribution but also the effective temperature of the radiation field, with higher values indicating the presence of smaller PAHs (and potentially higher temperatures); the 3.3/7.7 $\mu$m ratio can also trace ionization (A. G. G. M. Tielens 2008; B. T. Draine et al. 2021; D. Rigopoulou et al. 2021). The dependence of these ratios on PAH size and ionization stems from larger PAH sizes contributing proportionally more at longer wavelengths, while neutral PAHs contribute more at 3.3 and 11.3 $\mu$m than at 7.7 $\mu$m. Likewise, hotter radiation field temperatures (i.e., stronger heating) will shift power toward shorter wavelengths, producing an enhancement in the 3.3 $\mu$m feature, which is believed to be primarily produced by the smallest PAHs (B. T. Draine et al. 2021). Because collisions in the hot phase are expected to either rapidly erode small PAHs or remove them by incorporation onto larger dust grains (E. R. Micelotta et al. 2010a), tracking changes in these PAH band ratios provides insight into the state of the cooler phase of galactic winds.

### 4.1. General Trends

We present our PAH ratios maps in Figure 5, which are rotated to align the $y$-axis with the galaxy minor axis. The left panel features the 11.3/7.7 $\mu$m map. We find that variations in 11.3/7.7 $\mu$m occur on smaller scales, with perhaps a global trend of increasing ratio values with distance from the disk. We perform an average down a 33″-wide slice aligned with the minor axis (white rectangle). We then weight these ratios by the 11.3 $\mu$m intensity in order to emphasize regions with significant emission and suppress contributions from low-intensity regions where the ratios are poorly constrained. Weighting by intensity does not significantly change the results, and there is also consistency when using 7.7 $\mu$m or 3.3 $\mu$m intensity as the weighting scheme. These ratios are not corrected for distance to the starburst as in Figure 4. The result is a general trend of increasing 11.3/7.7 $\mu$m with distance from the starburst (blue line in left panel of Figure 6). These results are in good agreement with the observations by P. Beirão et al. (2015). In Appendix B, we apply synthetic photometry to mosaicked Spitzer IRS cubes of the M82 wind and disk from P. Beirão et al. (2008, 2015), and again find agreement in the overall

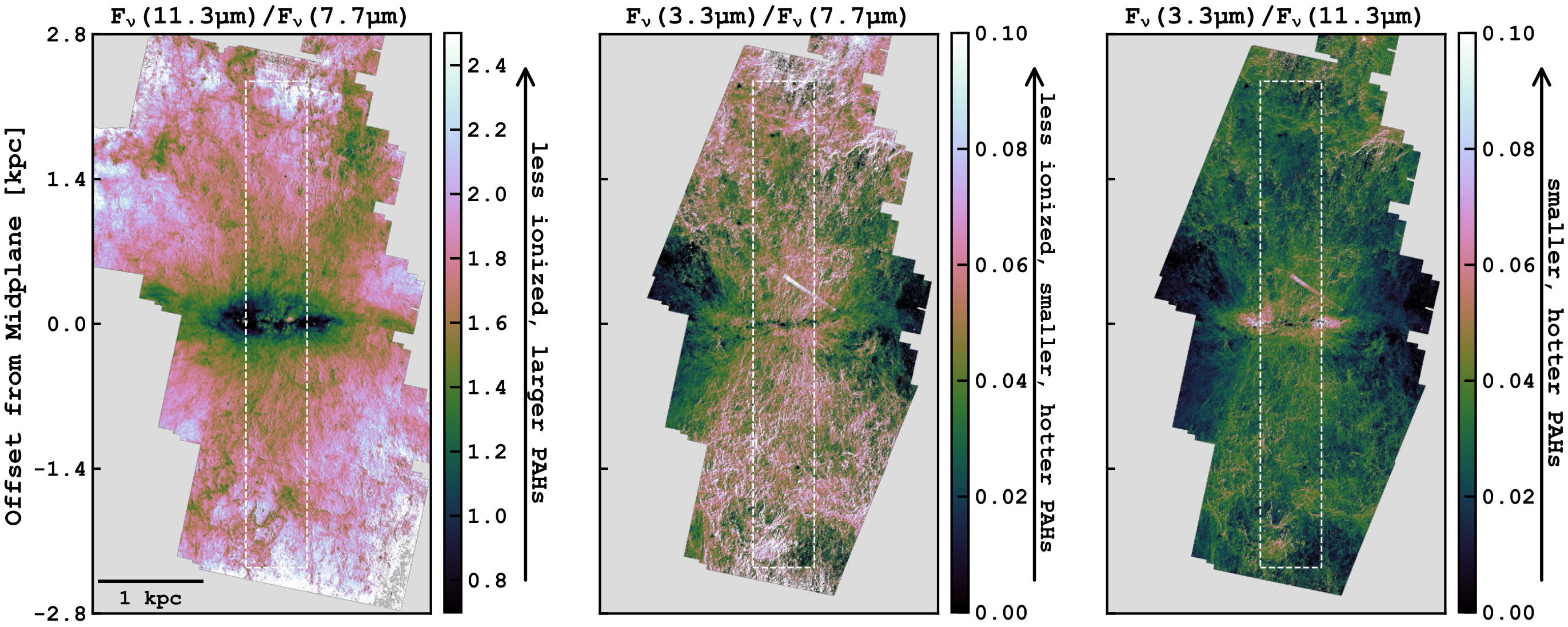

**Figure 5.** Ratios of PAH $F_\nu$ intensities [MJy sr$^{-1}$] in the M82 wind. Left panel: the 11.3/7.7 $\mu$m ratio, which is primarily sensitive to ionization. Larger ratio values indicate more neutral PAHs, but could also indicate larger PAHs. Middle panel: the 3.3/7.7 $\mu$m ratio. Larger ratio values primarily indicate smaller PAHs, though may hint at less ionization and hotter temperatures of the radiation field heating the PAHs. Right panel: similarly, the 3.3/11.3 $\mu$m ratio is most sensitive to size. Higher ratio values indicate smaller PAHs, and/or possibly hotter PAHs. In all three panels, the white rectangle represents the same 33″-width slice as in Figure 3. We plot the ratio trends over this slice in Figure 6. Both 3.3/7.7 and 3.3/11.3 $\mu$m ratios have an artifact near the center of the mosaic due to imprecise mosaicking of the SUB640 array with the FULL footprints. These ratio maps indicate that variations in PAH properties occur primarily on small scales, while global trends with distance from the starburst are less clear.

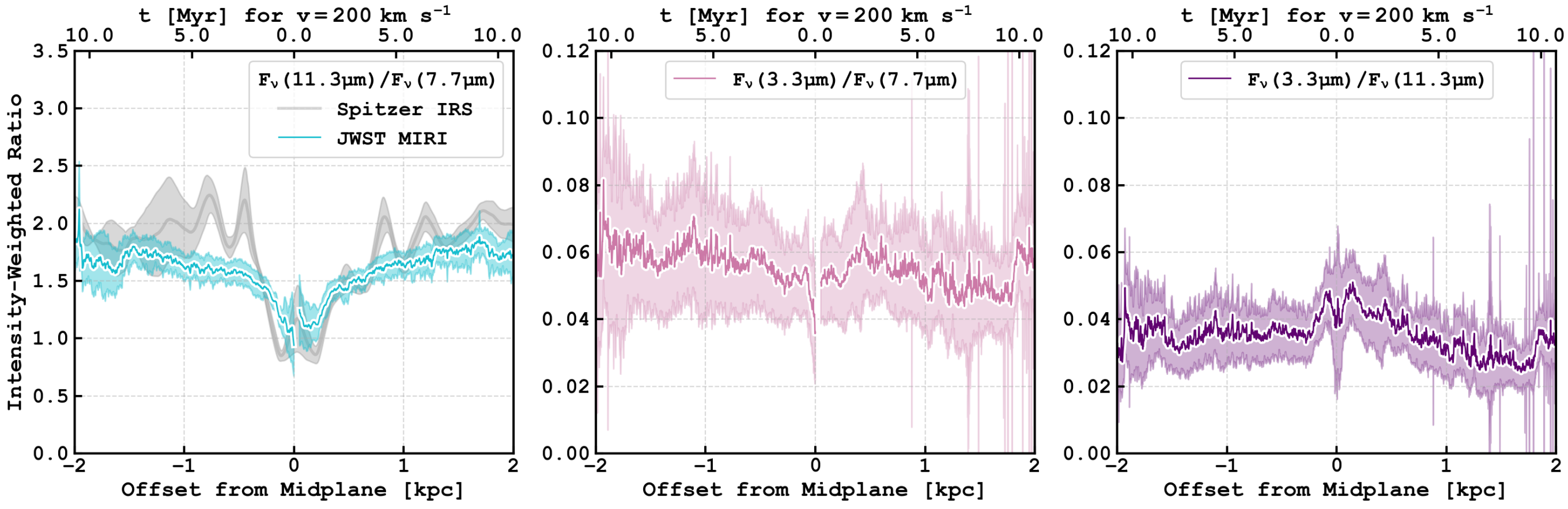

**Figure 6.** PAH ratios weighted by intensity at 11.3 $\mu$m as a function of distance in the M82 wind. These ratios are computed over the 33″-width slice in Figure 5, which matches the slice taken in Figure 4. An estimated timescale of material moving in the wind is on the top axis, assuming a deprojected velocity of 200 km s$^{-1}$. Left panel: the 11.3/7.7 $\mu$m ratio (blue) dips near the galaxy midplane, likely due to extinction due to the silicate feature at 9.7 $\mu$m affecting emission at 11.3 $\mu$m. There is a moderate gradient with distance that could be attributed to PAHs becoming more neutral with distance. These results are consistent with the 11.3/7.7 $\mu$m ratio measured with the IRS on Spitzer (gray; see Appendix B). Middle and right panels: the 3.3/7.7 and 3.3/11.3 $\mu$m ratios are mostly uniform with distance, an indicator of constant PAH size. The uptick in 3.3/11.3 $\mu$m in the nucleus is again likely due to silicate absorption.

11.3/7.7 $\mu$m trend between the Spitzer IRS data[32] (gray line in Figure 6) and JWST MIRI.

The middle and right panels of Figure 5 feature the 3.3/7.7 $\mu$m and 3.3/11.3 $\mu$m ratios, respectively, showing that variations in these ratios primarily exist on smaller scales and no obvious global trend with distance. Both ratios exhibit values an order of magnitude lower than 11.3/7.7 $\mu$m. The distribution of 3.3/7.7 and 3.3/11.3 $\mu$m, weighted by intensity and averaged over the 33″-width slice, can be found in the middle and right panels of Figure 6. The 3.3/7.7 and 3.3/11.3 $\mu$m ratios track each other well in the wind, with overall distributions appearing uniform within a factor of ∼2.

The flat 3.3/7.7 and 3.3/11.3 $\mu$m profiles suggest that, outside the nucleus, PAH size does not change significantly with distance from the starburst. This implies that PAHs could be well protected from shocks and hot gas out to ∼±2 kpc above and below the galaxy, perhaps due to environmental shielding by cold dense clouds. Alternatively, the small PAH population could be replenished by shock-induced grain–grain collisions and shattering (E. R. Micelotta et al. 2010b), although the flatness of the ratios would point to an almost perfect balance between these destruction and replenishment processes that we find implausible. We discuss these scenarios in detail in Section 6.

[32] The oscillations in the 11.3/7.7 $\mu$m ratio derived from Spitzer IRS are due to the PSF on the bright starburst (see Appendix B).

The 11.3/7.7 $\mu$m ratio is more difficult to interpret. The decrease in 11.3/7.7 $\mu$m toward the nucleus is likely dominated by extinction due to the broad silicate feature at 9.7 $\mu$m, which has been noted to affect PAH ratios in the M82 disk (P. Beirão et al. 2015), and is hard to correct for when estimating 11.3 $\mu$m using F1130W only (G. P. Donnelly et al. 2025b). This extinction effect is further evidenced by a slight increase in the 3.3/11.3 $\mu$m ratio in the nucleus and lack of change in 3.3/7.7 $\mu$m. As mentioned in Section 3, we estimate $\tau_{9.7} \approx 1.5$ in the disk. A $\tau \sim 1.2$ could decrease the 11.3/7.7 $\mu$m ratio by a factor of ∼2, according to the attenuation curves of luminous infrared galaxies drawn from the AKARI-Spitzer spectral dataset (T. S.-Y. Lai et al. 2024). Thus, silicate absorption plays a key role in the observed 11.3/7.7 $\mu$m ratio in the disk.

However, at distances beyond ±0.5 kpc, there appears to be a ∼10%–20% increase in 11.3/7.7 $\mu$m that may not be due to silicate absorption. Using a larger (100″-wide) rectangular slice or cone somewhat smooths the gradient, but a modest ∼5%–10% increase remains. This result could indicate that PAHs are becoming less ionized as they move farther away from the starburst and thus exposed to a declining radiation field. Other scenarios include PAHs growing in size by adding carbon atoms, or smaller PAHs could be selectively destroyed. However, without a corresponding increase in 3.3/7.7 $\mu$m (toward more neutral PAHs) and decrease in 3.3/11.3 $\mu$m (toward larger PAHs), it is difficult to interpret this 11.3/7.7 $\mu$m trend in isolation.

### 4.2. Comparison with Dust Models

To help interpret the observed PAH ratios, we compare our results with the dust models from B. T. Draine et al. (2021). To do so, we compute model grids by applying synthetic photometry on the B. T. Draine et al. (2021) model spectra, which depend on the assumed radiation field spectral shape and intensity. To match the absolute flux calibration of JWST, we use `synphot` to calculate the photon-weighted flux density transmitted through each photometric band (see Equation 5 of K. D. Gordon et al. 2022). Using the available JWST throughput curves (A. Glasse et al. 2015), we calculate synthetic photometric bands for F250M, F335M, F360M, F770W, and F1130W.[33] Passing the synthetic NIRCam bands through the algorithm from Section 2.6 results in a continuum-subtracted 3.3 $\mu$m PAH integrated intensity. We also subtract the continuum from the synthetic MIRI bands by assuming the median continuum contribution from Section 2.7.

After taking ratios between these modeled 3.3, 7.7, and 11.3 $\mu$m integrated intensities, the resulting grids can be found in Figure 7. We assume a G. Bruzual & S. Charlot (2003) starburst model for the radiation field with varying parameters. Testing variations in the starburst age and metallicity has determined that a 3 or 10 Myr starburst at solar metallicity (L. A. Lopez et al. 2020) works best. The full grids are displayed in the top row of the figure with zoomed-in versions below. Each panel contains three grids with different radiation field intensities: $U = 1, 10^2, 10^3$, informed by the radiation field from modeling dust emission calculated by A. K. Leroy et al. (2015). For readability, the points are the result of taking the mean of every $N = 70$ data points in Figure 6. The color bar represents vertical offset $z$ from the midplane. The arrows are informed by the model grids presented in B. T. Draine et al. (2021).

Generally, the ratios in both the northern and southern outflows are best explained with radiation field intensities of either $U = 1, 10^2, 10^3$. This is expected, as the spectral shape of the emission spectrum should not change with single-photon heating $\lesssim 15$ $\mu$m until $U \gtrsim 10^3$ (B. T. Draine et al. 2021; though we find no difference when we test against $U = 10^4$). However, S. Lopez et al. (2026) showed that vertical profiles of 8 $\mu$m PAH/$\Sigma_{\rm H\,I+H_2}$ (i.e., PAH-to-neutral-gas) and $\langle U \rangle$ (estimated from dust spectral energy distribution, SED, modeling of Herschel data; B. T. Draine & A. Li 2007; A. K. Leroy et al. 2015) appear to decline at the same rate over a distance of ∼2.5 kpc. The S. Lopez et al. (2026) profiles support the idea that the cool material as traced by PAHs is indeed affected by heating from the radiation field.

Comparison with the dust models (valid for $U = 1, 10^2, 10^3$) indicates that PAHs are of standard-to-large size (where the peak of the log-normal size distribution is 4 Å for "standard" and 5 Å for "large" PAHs; see Table 2 of B. T. Draine et al. 2021) and exhibit standard-to-high ionization. This result agrees with P. Beirão et al. (2015), who spectroscopically measured 11.3/7.7 and 6.2/7.7 $\mu$m ratios in the M82 wind with Spitzer IRS and compared them with dust models presented in B. T. Draine & A. Li (2001). In M82, we observe a gradient between standard and high ionization, with PAHs becoming more neutral with distance from the ionizing starburst. This result is consistent with D. A. Dale et al. (2023, 2025), who found that PAHs are more ionized in stellar clusters and stellar associations. Thus, cool material moving away from the source of the radiation field should be less susceptible to ionization.

We find a subtle trend of PAHs becoming larger with distance from the starburst in the northern wind (red points), while the southern wind shows a less clear trend (blue points). This contrast in PAH size between both wind lobes is evident in both the emission and ratio profiles (see Sections 3 and 4.1), and interestingly was also pointed out in P. Beirão et al. (2015). However, when we test averaging over (1) a larger, 100″-wide slice, and (2) a cone with a 30° semi-opening angle instead of a rectangular slice, this north–south difference is less discernible. The points at higher positive offsets move closer to the standard size distribution region of the grids, similar to the flattening of the surface brightness profiles when averaging over larger areas in the northern wind (Section 3). Therefore, results derived from the 33″ slice appear to also encode a low-emission region in the north, which is apparent in Figure 3, suggesting that the PAH size variations seen in Figure 7 may exist on smaller scales rather than globally and that azimuthal variations exist across the wind. Even though these variations occur mostly on small scales, it is interesting that they are most prominent in the northern lobe of the wind. This dichotomy could reflect differences in the ambient media into which the outflow expands. For instance, the southern lobe of the M82 outflow is known to be interacting tidally with M81, leading to an asymmetry in H I extent between the southern and northern outflows (P. Martini et al. 2018). This tidal interaction could alter the density and turbulent structures of the surrounding CGM, providing different environmental conditions that PAHs are exposed to. S. Lopez et al. (2025) also supplied this as a possible reason for more frequent prominent H$\alpha$-arcs in the south versus the north. Morphological differences between the north

[33] Note that the synthetic bands in B. T. Draine et al. (2021) ($F_{\rm clip}$) are calculated over wavelength ranges different from the JWST throughput filters used here.

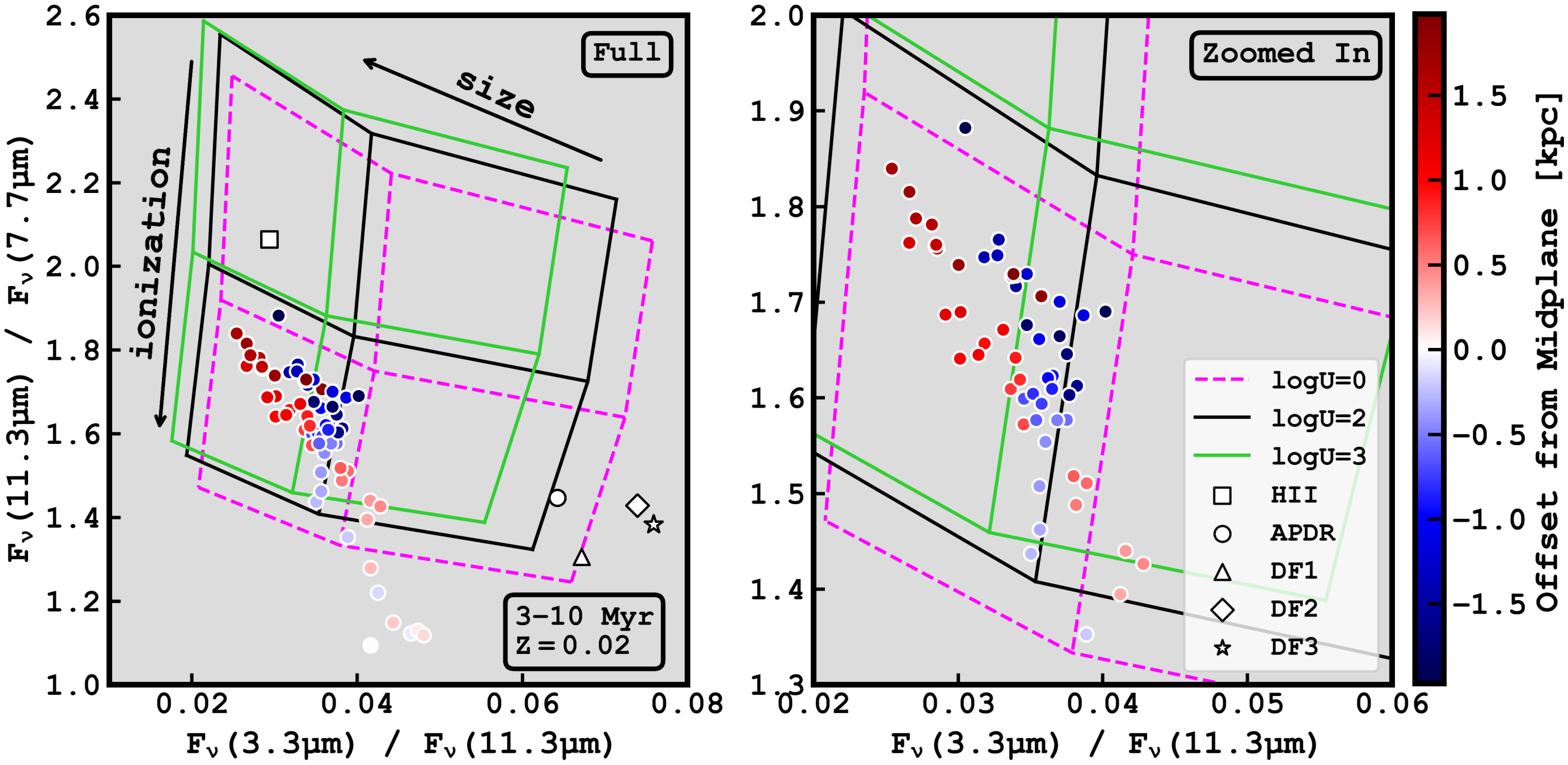

**Figure 7.** M82 PAH intensity-weighted ratios plotted over synthetic ratio grids calculated from the B. T. Draine et al. (2021) dust models. The points represent the mean ratio after binning the Figure 6 data for readability. The right panel is a zoomed-in version of the left panel to easily distinguish the points. The grids show very little change when assuming a G. Bruzual & S. Charlot (2003) radiation field from either a 3 or 10 Myr old starburst at solar ($Z = 0.02$) metallicity. The individual grids vary only in the mean radiation field heating of the dust ($U$). The choice of $U = 1$, $10^2$, $10^3$ is informed by independent derivations of the M82 radiation field intensity from B. T. Draine & A. Li (2007) spectral energy distribution (SED) fitting to Herschel data (A. K. Leroy et al. 2015), and from SOFIA observations (R. C. Levy et al. 2023). The ratios found in the starburst are not well represented by any of the models, including the solar neighborhood or M31 bulge models (not shown), and are likely driven by silicate absorption diluting the 11.3 $\mu$m intensity. Overall, PAHs in the M82 wind range between standard-to-high ionization state and standard-to-large size. PAHs appear more neutral (and possibly larger) with distance, with charge driving most of the observed trend. The white shapes are ratios measured in the H II region, atomic PDR (APDR), and dissociation fronts (DFs) of the Orion Bar (E. Peeters et al. 2024; R. Chown et al. 2024). The M82 ratios match closest with values measured in the H II region of the Bar.

and the south also exist in the X-ray outflow (D. K. Strickland et al. 1997).

The observed ratios in the central starburst region fall outside the grids toward higher ionization. Neither the solar neighborhood nor the M31 bulge models from B. T. Draine et al. (2021) explain these values (as they are very low-energy environments compared to M82), and using the starburst models with even higher $U$ does not alleviate this issue. Extinction near the starburst almost certainly plays a role, with silicate absorption affecting the 11.3 $\mu$m complex as discussed in Section 3. Additionally, the central starburst is a complex region that is nearly edge-on ($i \sim 80°$; Y. D. Mayya et al. 2005) with overlapping lines of sight and considerable dust attenuation even in the near-IR ($A_J \sim$ 1–7 mag; R. C. Levy et al. 2024). It is therefore difficult to interpret the ratios measured in M82's disk.

Finally, for comparison, we include data points on the Orion Bar from PDRs4All (white shapes in Figure 7; O. Berné et al. 2022). To obtain these points, we perform synthetic photometry on the template MIRI (R. Chown et al. 2024) and NIRSpec (E. Peeters et al. 2024) spectra for five distinct regions in the photodissociation region (PDR): the H II region, the atomic PDR (APDR), and three dissociation fronts (DFs) in the molecular PDR. We follow the same `synphot` procedure outlined earlier in this section to obtain synthetic F250M, F335M, F360M, F770W, and F1130W photometry in these regions. To subtract the continuum, we use the A. D. Bolatto et al. (2024) prescription to isolate 3.3 $\mu$m PAH emission; for the MIRI bands, we remove the fractional continuum contribution in each region that was estimated using synthetic photometry in R. Chown et al. (2025). We find that applying the latter method to F335M results in negligible differences in the estimated 3.3 $\mu$m PAH emission compared to the A. D. Bolatto et al. (2024) method. The numerical differences between this work and the ratios derived by the PDRs4All team (A. Maragkoudakis et al. 2026) stem from differing integration methods on the spectroscopy. The overall trends between each Orion Bar region shown in Figure 7 match what is found by PDRs4All.

We find a clear shift between the M82 ratio values and the Orion Bar. Most notably, 3.3/11.3 $\mu$m is higher in the APDR and DF regions than M82. This indicates that the molecular and atomic regions of the Orion Bar are populated with smaller PAHs than in the M82 wind. However, PAHs appear larger in the H II region, matching well with M82. It is possible that preferential destruction of smaller PAHs is at play here, with the denser DFs and APDR providing favorable conditions to protect the smallest PAHs, whereas the more diffuse M82 wind and Orion H II region are less hospitable. The ionization state of PAHs (11.3/7.7 $\mu$m) in the atomic and molecular regions of the Bar matches that of M82 more closely than the inferred PAH sizes, though PAHs in the Bar are on average still more ionized than those in the M82 wind. The exception is the H II region of the Bar, which are more neutral than those in the M82 wind. Notably, as the PDR is face-on, the H II region lies in front of the background molecular cloud. This means that the observed PAH emission primarily arises from PAHs at the surface of the cloud. Because the M82 wind ratios are closest to the H II region of the Bar, it is possible that the PAHs we observe in the wind also arise from the surface layers of cool clouds. We further discuss this idea in Section 6.2.

## 5. PAH Abundance

Another indicator of cool-material processing, as traced by PAHs, is the PAH abundance relative to the total dust content in the wind, $q_{\rm PAH}$. More precisely, $q_{\rm PAH}$ is the fraction of mass in the form of PAHs with $<10^3$ carbon atoms relative to the total dust mass (B. T. Draine & A. Li 2007; B. T. Draine et al. 2021). For reference, $q_{\rm PAH}$ is typically ∼5% for solar-metallicity spiral galaxies (B. T. Draine & A. Li 2007; B. T. Draine et al. 2007). Although $q_{\rm PAH}$ is not a directly observable quantity, one can approximate it by fitting dust emission models to the full infrared SED (e.g., B. T. Draine et al. 2007), which simultaneously constrains the radiation field intensity $U$. Without full SED-fitting, a $q_{\rm PAH}$ proxy can be obtained from the ratio of total PAH luminosity ($L_{\rm PAH}$) to the total IR luminosity ($L_{\rm TIR}$), assuming $U$. Using Spitzer IRS mapping of three nearby spiral galaxies, combined with SED-fitting using IR photometry from Spitzer and the Herschel Space Observatory, C. M. Whitcomb et al. (2024) found that $L_{\rm PAH}/L_{\rm TIR}$ is tightly correlated with $q_{\rm PAH}$. This correlation holds true in the low radiation field limit ($U \lesssim 10^3$) but becomes uncertain in more extreme environments such as starbursts (B. T. Draine & A. Li 2007); however, this should be a good approximation in the M82 wind given $U \lesssim 10^2$ at vertical heights $z > 0.5$ kpc (A. K. Leroy et al. 2015).

We define $L_{\rm PAH} = 2 \times L_{7.7\ \mu{\rm m}}$, assuming that the 7.7 $\mu$m PAH complex contributes up to 50% the total PAH power (Figure 9 of J. D. T. Smith et al. 2007). Hereafter, we refer to this quantity as $L_{\rm PAH}$(JWST) (this is not corrected for the radiation field as in Figure 4). We compute $L_{7.7\ \mu{\rm m}} = \Delta\nu L_\nu(7.7\ \mu{\rm m})$, converting the spectral luminosity to an integrated luminosity using a $\Delta\nu$ corresponding to an assumed bandwidth of 1 $\mu$m for the 7.7 $\mu$m complex (E. Peeters et al. 2002; J. D. T. Smith et al. 2007).

With the limited coverage offered by the JWST data, we can only probe $q_{\rm PAH}$ out to distances of ±2.5 kpc, and timescales of ∼10 Myr (assuming $v \approx 200$ km s$^{-1}$). To capture PAH abundance on longer timescales (∼20 Myr), where destruction and growth processes may be more relevant, we also perform this exercise using Spitzer IRAC4 8 $\mu$m imaging (Figure 1). We first apply color- and extended-source corrections to the Spitzer image and convert to the same photometric convention of the F770W image (see Appendix C for details). After subtracting a continuum contribution of ∼16% (our estimate from Section 2.7), the 8 $\mu$m image is dominated by aromatic PAH features, though it may also contain other dust emission much like the MIRI images (C. W. Engelbracht et al. 2006). While the IRAC4 channel is centered on 8 $\mu$m, it captures the 7.7 $\mu$m PAH complex similar to F770W. To estimate the total PAH luminosity, we define $L_{\rm PAH}$(Spitzer) $= 2 \times L_{8\ \mu{\rm m}}$, where $L_{8\ \mu{\rm m}} = \Delta\nu L_\nu(8\ \mu{\rm m})$ (again for a bandwidth corresponding to 1 $\mu$m), matching the $L_{\rm PAH}$(JWST) definition above.[34]

To determine $L_{\rm TIR}$, we use archival photometry from the PACS and SPIRE instruments on board Herschel (H. Roussel et al. 2010). We adopt the M. Galametz et al. (2013) calibrations to combine the Herschel bands into actual measurements of $L_{\rm TIR}$ (see their Tables 2 and 3 for the calibration coefficients). M. Galametz et al. (2013) found that 70 $\mu$m emission is a reasonable monochromatic estimate of $L_{\rm TIR}$, and a multiband combination using 70 + 160 $\mu$m provides an estimate of $L_{\rm TIR}$ with a precision of 25%. While a three-band calibration could improve this precision, the only other Herschel observations available are at wavelengths >250 $\mu$m, which would result in a severe loss of angular resolution ($\gtrsim$18″; G. Aniano et al. 2011). This would lead to a high degree of contamination from the very bright starburst into the wind, which is the object of our study.

We convolve the Herschel and Spitzer maps to the PSF of the longest wavelength (∼6″ at 70 $\mu$m and ∼12″ at 160 $\mu$m) using the G. Aniano et al. (2011) convolution kernels; for $L_{\rm PAH}$(JWST), we convolve the JWST maps to the Herschel PSFs also presented in G. Aniano et al. (2011). To avoid contaminating the faint wind with emission from the bright starburst, we exclude data out to a radius of 1 kpc from the midplane. We test excluding data less than and greater than this radius, and adopt 1 kpc as a compromise that limits starburst contamination while retaining at least half of the wind area. We then calculate $q_{\rm PAH}$ [%] $= (L_{\rm PAH}/L_{\rm TIR}$ [%])/(3.69 ± 0.02) for $L_{\rm TIR}$(70 $\mu$m) and $L_{\rm TIR}$(70 + 160 $\mu$m), where the multiplicative factor was derived in C. M. Whitcomb et al. (2024) by performing a linear fit to the positive correlation between ΣPAH/TIR (i.e., ($L_{\rm PAH}/L_{\rm TIR}$)) and $q_{\rm PAH}$ found in varying environments across a sample of nearby galaxies.

There are two main caveats to this approach. First, while M. Galametz et al. (2013) calibrated $L_{\rm TIR}$ on local ($d \lesssim 30$ Mpc) galaxies that span a wide range of morphologies, star formation activities, and metallicities, the calibrations primarily account for dust emission in the main disks of galaxies rather than extraplanar wind material. Dust in winds may be subject to different heating sources, grain processing, and physical conditions than galaxy disks, which could alter the derived $L_{\rm TIR}$ values. Similarly, the conversion from $L_{\rm PAH}/L_{\rm TIR}$ to $q_{\rm PAH}$ in C. M. Whitcomb et al. (2024) was also only determined for a handful of galaxies and regions, and so may not be a perfect calibration for M82. Second, we find that the use of monochromatic versus combined tracers of $L_{\rm TIR}$ produce similar trends but are offset by a factor of ∼2. These differences arise because the 70 and 160 $\mu$m bands probe different parts of the dust temperature distribution, meaning monochromatic calibrations introduce temperature-dependent uncertainties that are mitigated when using multiband calibrations. Indeed, we find the 70 $\mu$m/160 $\mu$m ratio in M82 and its wind corresponds to dust temperatures of ∼20–60 K. This is in line with the 12–50 K calculated by H. Roussel et al. (2010) using ratios of 250 $\mu$m/350 $\mu$m and 250 $\mu$m/500 $\mu$m. These temperatures also decrease with distance from the starburst. Additionally, M. Galametz et al. (2013) acknowledged that using solely 70 $\mu$m emission can overpredict $L_{\rm TIR}$ in objects with warmer dust temperatures. As a consequence of these caveats, we treat our computation of $L_{\rm TIR}$ in the wind as an approximation and focus on relative trends rather than absolute numbers for $q_{\rm PAH}$.

[34] We also test definitions of $L_{\rm PAH}$(JWST) $= L_{7.7\ \mu{\rm m}} + L_{11.3\ \mu{\rm m}} + L_{3.3\ \mu{\rm m}}$ and $L_{\rm PAH}$(Spitzer) $= 1.2 \times L_{8\ \mu{\rm m}}$ rather than assuming the 7.7 $\mu$m complex makes up ∼50% the total PAH power. This approach utilizes all of the PAH bands at our disposal, adopting bandwidths from E. Peeters et al. (2002) and J. D. T. Smith et al. (2007). In this case, the 11.3 $\mu$m and 3.3 $\mu$m luminosities contribute ∼20% of $L_{\rm PAH}$(JWST), which is a reasonable estimate (J. D. T. Smith et al. 2007). However, this omits key contributors such as the 6.2 $\mu$m feature, which accounts for ∼15% of the total PAH power. These definitions reduce our $q_{\rm PAH}$ estimate to $\lesssim$1%, but do not change our conclusions.

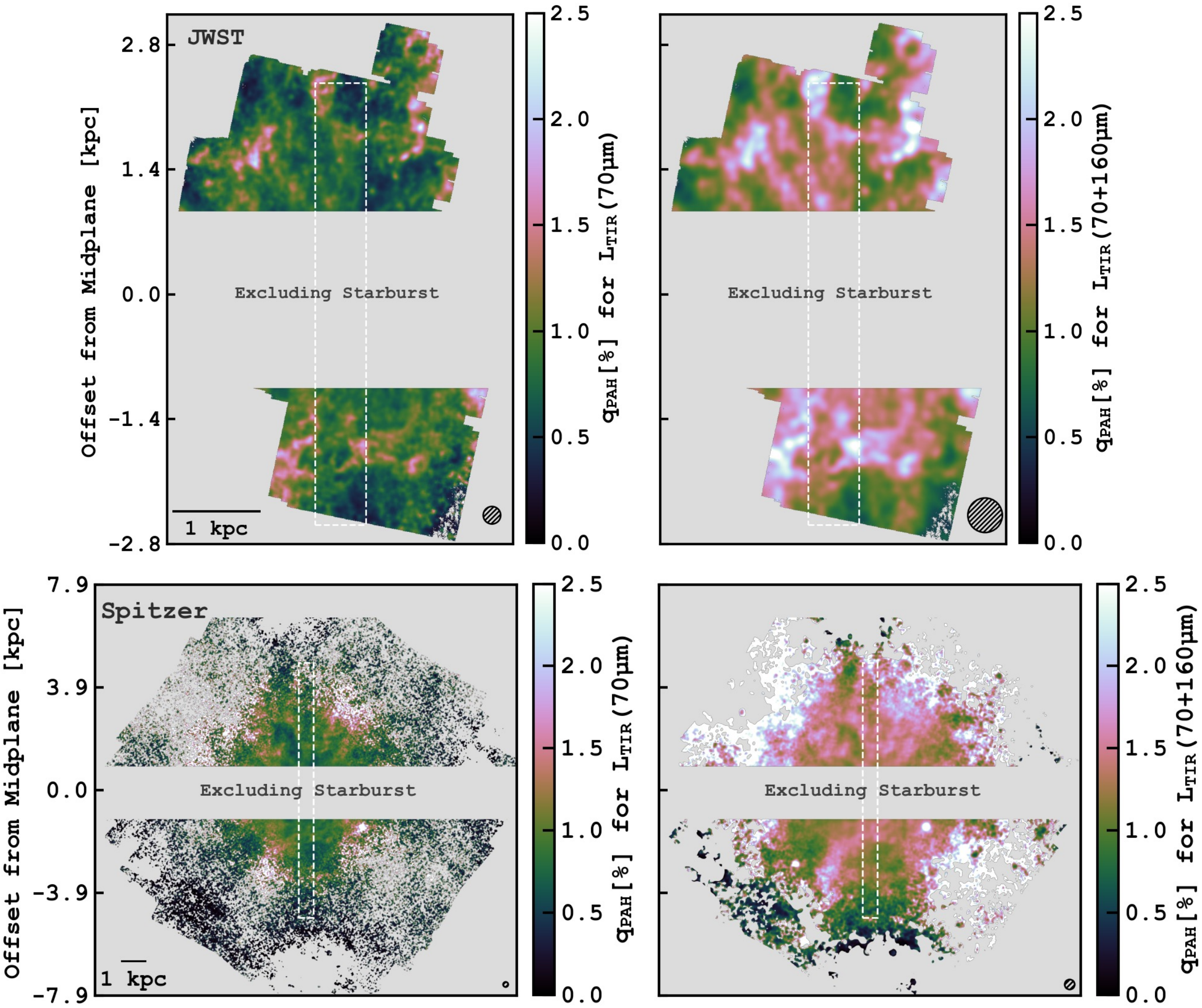

**Figure 8.** $q_{\rm PAH}$ calculated as $q_{\rm PAH}$ [%] $=(L_{\rm PAH}/L_{\rm TIR}$ [%])/(3.69 ± 0.02) (C. M. Whitcomb et al. 2024). We test between two M. Galametz et al. (2013) calibrations for $L_{\rm TIR}$: 70 $\mu$m and 70 + 160 $\mu$m, both data taken by Herschel. The resulting numerical difference between the two calibrations is the result of the different dust temperatures probed at these wavelengths. To avoid effects from the large Herschel beams (bottom-right corners; ∼6″ at 70 $\mu$m and ∼12″ at 160 $\mu$m), we exclude the starburst by removing data within ±1 kpc of the galaxy midplane. The top panels estimate the PAH luminosity as a 2 × $L_{7.7\mu{\rm m}}$ ($L_{\rm PAH}$(JWST)), assuming the 7.7 $\mu$m feature accounts for ∼50% of the total PAH power. The bottom panels represent the same exercise but instead using Spitzer 8 $\mu$m ($L_{\rm PAH}$(Spitzer)) to probe $q_{\rm PAH}$ at larger distances and, thus, longer timescales. $L_{\rm PAH}$(Spitzer) contains aromatic features that are mostly due to PAHs in the 7.7 $\mu$m complex (C. W. Engelbracht et al. 2006). Regions of more elevated $q_{\rm PAH}$ correspond with filaments where gas is likely dense and cool enough to protect PAHs.

The resulting ratio maps are presented in Figure 8, with the top panels using $L_{\rm PAH}$(JWST) and the bottom panels using $L_{\rm PAH}$(Spitzer). The leftmost maps use $L_{\rm TIR}$(70 $\mu$m), and the rightmost maps use $L_{\rm TIR}$(70 + 160 $\mu$m). While smoothed to the larger Herschel beams, the JWST maps still show some structure in the wind, and $q_{\rm PAH}$ appears slightly elevated in the wind filaments. These filamentary structures are likely denser and cooler than the surrounding medium, making them more conducive to shielding PAHs. Using the Spitzer map, $q_{\rm PAH}$ appears globally uniform on larger scales.

Again, we average these $q_{\rm PAH}$ values down the white slice in Figure 8 and plot the trends in Figure 9. These averages are weighted by intensity, so low-emission pixels are down-weighted. Including the Spitzer map allows us to sample $q_{\rm PAH}$ out to ±5 kpc from the midplane, which, at the typical wind speed of several hundred kilometers per second, corresponds to a few million years—more than enough time to destroy PAHs embedded in the hot wind (where $\tau_{\rm sputter}$ ∼ few thousand years; E. R. Micelotta et al. 2010a).

All four estimates of $q_{\rm PAH}$ show an overall uniform profile with distance from the midplane, remaining within a factor of 2 and the median hovering around ∼1%. This is less than the typical $q_{\rm PAH}$ found in the solar-metallicity ISM (∼5%; B. T. Draine & A. Li 2007; B. T. Draine et al. 2007). We interpret this result as PAHs being destroyed in the starburst due to the intense radiation field and shocks. Because M82 is at solar metallicity, the low $q_{\rm PAH}$ cannot be attributed to metallicity effects. Instead, it is consistent with the suppression of $q_{\rm PAH}$ in environments exposed to strong radiation fields such as H II regions (A. Topchieva et al. 2018; O. V. Egorov et al. 2023, 2025). Outside the starburst, the flat profiles suggest that the PAHs that survive long enough to be launched in the wind are unchanged on a timescale of at least ∼20 Myr. This result could indicate that destruction and growth processes of PAHs are in equilibrium with each other in the wind. The more likely scenario is that that PAHs are being safely transported over timescales that would otherwise ensure destruction without environmental shielding. We discuss these competing hypotheses at length in the next section.

Given the caveats listed earlier in this section, the above calculation may be more appropriate for a constant $U \sim 1$. To see how our results may be affected by this assumption, we test a derivation of $q_{PAH}$ that accounts for the changing radiation field in the M82 wind. To do so, we calculate $L_{TIR}$ from Figure 15 in B. T. Draine & A. Li (2007), which plots $\langle \nu L_\nu(70\ \mu m)\rangle / L_{TIR}$ and $\langle \nu L_\nu(160\ \mu m)\rangle / L_{TIR}$ as a function of $U$ for various percentages of $q_{PAH}$. Here, we use $U$ derived in A. K. Leroy et al. (2015) to perform the calculation. We then convert to $q_{PAH}$ using the equation from C. M. Whitcomb et al. (2024) for both $L_{PAH}$(JWST) and $L_{PAH}$(Spitzer). We find no meaningful difference between the estimate of $q_{PAH}$ using $L_{TIR}$ derived from $\nu L_\nu(70\ \mu m)$ and our results above, with the median remaining ∼1%. There is an increase in $q_{PAH}$ when using the $L_{TIR}$ derived from $\nu L_\nu(160\ \mu m)$, with a median at ∼2%, but this does not change our conclusions.

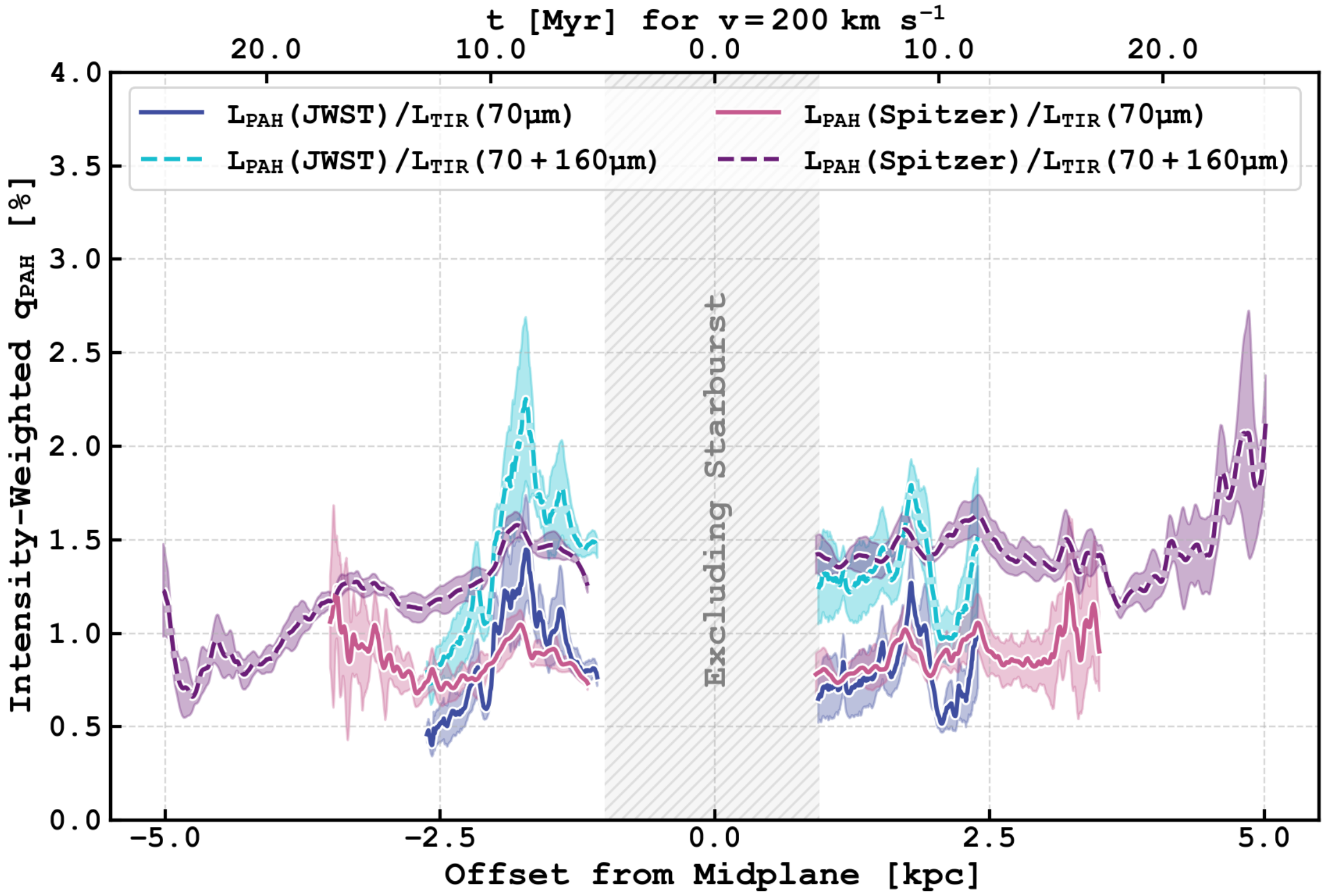

**Figure 9.** $q_{PAH}$ profiles with respect to distance from the midplane, averaged over the 33″-wide slices shown in Figure 8. The profiles are intensity-weighted before being averaged in order to down-weight pixels with low emission. The shaded regions represent the standard deviation of the values being averaged. Within a factor of ∼2, we find little variation in $q_{PAH}$ with vertical distance out to ±5 kpc. The median $q_{PAH}$ in the M82 wind is ∼1%, below typical values for the Milky Way and solar-metallicity disk galaxies (∼5%; B. T. Draine et al. 2007; B. T. Draine & A. Li 2007).

## 6. Discussion

The goal of this work is to track the evolving state of entrained cool material in the M82 wind as it is transported toward the CGM. PAHs provide a unique window into the physical state of cool material, as hotter phases of galactic winds provide conditions that are inhospitable to the long-term survival of PAHs (E. R. Micelotta et al. 2010a, 2010b). In M82, variations in the observed PAH luminosity appear to be governed by distance to the illuminating starburst (Figure 4). Comparison with dust models reveals that the M82 wind is populated with large and ionized PAHs (Figure 7). The 3.3/11.3 and 3.3/7.7 μm ratios show very little change over the course of ∼10 Myr, whereas 11.3/7.7 μm may show a slight increase outside the starburst. Finally, there is no strong evolution in the PAH abundance $q_{PAH}$ over the course of ∼20 Myr (Figure 9). These timescales are relevant for our assumed outflow velocity of 200 km s$^{-1}$ (see discussion in Section 3).

### 6.1. Influences on PAH Band Ratios

PAHs being larger and more ionized in the M82 wind is supported by Spitzer spectroscopic results (P. Beirão et al. 2015) and by known correlations between grain size and charge. Calculations of photoelectric heating by dust grains have determined that, for nanoparticles, the fraction of grains that are charged depends on size (J. C. Weingartner & B. T. Draine 2001a; B. S. Hensley & B. T. Draine 2017). Focused specifically on PAHs, laboratory experiments have demonstrated that larger PAHs are more likely to undergo ionization than fragmentation when exposed to harsh UV radiation (J. Zhen et al. 2015, 2016; G. Wenzel et al. 2020), which is also supported by theoretical calculations (T. Allain et al. 1996a). Supporting this idea, spectral modeling of several hundred PAH molecules found that the observed PAH band ratios in M82 are predominantly produced by cationic, large PAHs of about ∼80–100 carbon atoms (A. Maragkoudakis et al. 2020). When mapping our 3.3/11.3 μm ratio to PAH intensity versus size models of A. K. Lemmens et al. (2023), the wind may be populated with PAHs containing ∼100 or more carbon atoms.

In M82, the overall flat 3.3/11.3 and 3.3/7.7 μm ratios indicate very little change in PAH size with distance from the galaxy, though a marginal trend emerges when plotting them against the B. T. Draine et al. (2021) dust models (Figure 7); this trend becomes smeared and less significant when measured over larger, more global scales (see Section 4.2).

This near-constancy of the PAH size distribution with distance may be maintained by a cycle of coagulation and shattering. PAHs coagulating onto larger grains removes them from the ISM through a nondestructive channel. Moderate ($\sim$1–2 km s$^{-1}$) grain–grain collisions, driven by magnetohydrodynamic turbulence in dense molecular clouds, then shatters these large grains into smaller fragments, restoring this PAH material to the nanoparticle population (H. Yan et al. 2004). This recycling of material from larger grains provides a continuous source of small PAHs that could regulate the size distribution. In contrast, the Makani Galaxy exhibits a decreasing $(11.3 + 12.2\ \mu\mathrm{m})/7.7\ \mu\mathrm{m}$ ratio with distance from the galaxy nucleus, suggesting that PAHs are eroding and becoming more ionized as they are transported in the Makani wind (S. Veilleux et al. 2025). This difference may reflect the longer timescales over which PAHs are exposed to in Makani ($\sim$100 Myr) compared to the region currently probed in M82 ($\sim$10–20 Myr).

More significant changes in the observed PAH band ratios in M82 appear to be dominated by charge, with PAHs becoming more neutral with distance from the ionizing starburst (Figure 7). Spatially resolved MIRI-MRS and NIRSpec-IFU spectroscopy of galaxies also determined that PAH charge influences the PAH band ratios more than grain size, extinction (by ices or silicates), and dehydrogenation (D. Rigopoulou et al. 2024). If charge is indeed the dominant factor in shaping the PAH band ratios in M82, then variations in the ionization fraction must produce the 11.3/7.7 and 3.3/11.3 $\mu$m trends with distance observed in Figures 6 and 7.

Both theory and observations point toward radiation field strength and hardness as drivers of variations in PAH ionization. In the case of M82, this means a changing radiation field could be the culprit behind an evolution of PAH charge with distance from the central starburst. In particular, stronger UV radiation and the spectral shape of the radiation field may alter the PAH ionization fraction (A. Omont 1986; V. Le Page et al. 2001; B. T. Draine et al. 2021; D. Rigopoulou et al. 2021, 2024). After absorbing UV photons, ionization and fragmentation compete depending on PAH size and internal energy. In larger PAHs, ionization often dominates over fragmentation (T. Allain et al. 1996b). Within this context, large PAHs in the M82 wind may be ionized near the starburst without also undergoing fragmentation. As PAHs are transported away from the starburst, they are then exposed to a weaker and/or softer radiation field, which declines by $\sim$2 orders of magnitude between $z \approx 0$ and 2.5 kpc (A. K. Leroy et al. 2015; R. C. Levy et al. 2023) and coincides with decreasing PAH intensity over this same vertical height (S. Lopez et al. 2026). In terms of the ionization parameter ($\gamma = G_0 T^{1/2}/n_e$; E. L. O. Bakes & A. G. G. M. Tielens 1994), this implies a decrease in $\gamma$ with distance from the starburst. We assume $G_0$ (the strength of the FUV field in the 6–13.6 eV range, normalized to the Habing field, i.e., the typical radiation field in the solar neighborhood) declines as $\sim 10^{3.5}$ in the starburst to $\sim 10^{1.5}$ in the wind (R. C. Levy et al. 2023), and that the electron density decreases from $n_e \sim$ 10–30 cm$^{-3}$ to $n_e \sim$ 1–10 cm$^{-3}$ over the same distance (typical for diffuse galactic winds; M. S. Westmoquette et al. 2009; S. A. Cronin et al. 2025), assuming pressure equilibrium between the surrounding ionized gas and the PDR. Taking a temperature of $T \sim 100$ K (P. Beirão et al. 2015), characteristic of PDR gas where PAHs reside, we then expect the ionization parameter $\gamma$ to decline from $\sim 10^3$ near the disk to $\sim 10^2$ in the outer wind. The weakening radiation field and declining ionization parameter thus produce the observed increase in 11.3/7.7 $\mu$m ratio with distance (Figure 6). In practice, this would also generate a subsequent increase in 3.3/7.7 $\mu$m, which we do not observe (though PAH size may also play some role in shaping this ratio). The ionization parameter derived for M82 is less than that of the Orion Bar ($\gamma \sim 4 \times 10^4$; A. Sidhu et al. 2022), in line with Figure 7, which shows that PAHs in the Bar are on average more ionized than the PAHs in the M82 wind.

D. Baron et al. (2024, 2025) offered an alternative perspective: the radiation field may directly impact band ratios through PAH heating rather than an indirect approach through modifying PAH size or charge. Implementing machine learning techniques on 40–150 pc regions across 19 galaxies in the PHANGS sample, these papers linked changes in both the 11.3/7.7 and 3.3/11.3 $\mu$m ratios with varying radiation field, supporting previous claims from dust modeling (B. T. Draine et al. 2021). This scenario assumes PAH charge and size remain constant and that the radiation field directly alters PAH band ratios by heating PAHs to higher internal temperatures and shifting their emission to shorter wavelengths (A. G. G. M. Tielens 2008; B. T. Draine et al. 2021). In this framework, the uptick in 3.3/11.3 $\mu$m toward the M82 nucleus could be a consequence of the starburst radiation field in addition to silicate absorption. Only after accounting for the radiation field did D. Baron et al. (2024) recover the connection between 3.3/11.3 $\mu$m and the PAH size distribution in these galaxies. Likewise, the 11.3/7.7 $\mu$m ratio would reflect a varying radiation field instead of significant changes in PAH ionization.

While a nonvarying charge and size distribution may oversimplify reality (and considering that these studies did not observe extraplanar outflows), the conclusion that PAH band ratios are directly influenced by the radiation field is supported by simulations. Galaxy evolution models emphasize that the leading contribution to PAH band intensity (and thus variations in band ratios) is the radiation field SED rather than, e.g., the grain size distribution (D. Narayanan et al. 2023). Under the single-photon approximation, which would be a reasonable assumption outside the dense starburst of M82, dust modeling done by H. M. Richie & B. S. Hensley (2025) supported a degeneracy between physical PAH properties and the underlying radiation field spectral shape and intensity (see their Figure 6). The 3.3 $\mu$m feature has a particular sensitivity to the hardness of the radiation field. Overall, these results point toward hardness of the radiation field controlling the conditions of the cool phase of the M82 wind, either by directly setting the PAH band ratios or indirectly influencing them by impacting PAH size and charge.

Thus far, we have assumed that UV-optical photons from the starburst are the primary heating source of PAHs in the wind. Dust models from, e.g., B. T. Draine et al. (2021), and D. Rigopoulou et al. (2021) focus on PAH heating from UV-optical radiation fields and do not explicitly model the effects from X-rays. Indeed, GALEX observations of the M82 wind show that a significant fraction of UV light escapes the disk and is scattered off dust, creating an extended UV radiation field in the outflow that reaches at least $\sim$6 kpc beyond the disk (C. G. Hoopes et al. 2005; C. T. Coker et al. 2013), similar to the extent of the PAH emission. While the radiation pressure is insufficient to drive the outflow (as indicated by a low gas Eddington ratio; C. T. Coker et al. 2013), the direct UV photons from the radiation field provide sufficient energy to heat PAHs in the wind. However, galactic winds driven by nuclear starbursts are typically multiphase cones filled with $T \sim 10^7$ K, X-ray-emitting gas that is cocooned by a wall of entrained, cooler material (e.g., T. M. Heckman et al. 1990; D. D. Nguyen & T. A. Thompson 2021; T. A. Thompson & T. M. Heckman 2024). This raises the question of whether heating by X-ray photons from the hot phase of the wind contributes to the observed PAH emission or if UV radiation remains the dominant heating source.

Laboratory experiments have shown that absorbing 0.2–2.5 keV X-rays leads to multiply ionized PAHs (i.e., PAHs that have lost multiple electrons), making them vulnerable to fragmentation, and whose half-lives are too short to account for the observed IR emission (T. Monfredini et al. 2019; Y. Huo et al. 2023a, 2023b). However, these studies only considered PAHs of $N_c \sim 10$–24 carbon atoms. As discussed earlier in this section, the PAHs launched in the M82 wind could be larger, and larger PAHs are generally more resistant to fragmentation (T. Allain et al. 1996b; Y. Huo et al. 2023a). Moreover, environmental shielding by cold gas and grain growth may allow PAHs to survive under harsh X-ray conditions (T. Monfredini et al. 2019). The calculations of E. R. Micelotta et al. (2010a) also found that, when exposed to the $T \gtrsim 10^6$ K wind, collisions with electrons play a larger role in PAH destruction than the absorption of X-ray photons. However, a key difference in the case of X-rays is their ability to penetrate into the neutral layer of PDRs where UV photons are absorbed. X-ray flux may be too low to contribute to PAH heating, and instead lead to single-photon fragmentation of PAHs with 50 carbon atoms, whereas larger PAHs may require multiple X-ray photons (K. Lange et al. 2025). In essence, if X-ray photons do indeed contribute to the heating and ionization of large PAHs in the wind, then comparisons with dust models like in Figure 7 offer an incomplete picture, since we only focus on the effects from the UV-optical radiation field. Modeling the full effects of X-ray photons on entrained PAH molecules in galactic winds could be an interesting avenue for future study.

### 6.2. What Is Driving the PAH Abundance?

PAH abundance, $q_{\rm PAH}$, is regulated by a balance between PAH formation and destruction processes in the ISM. This balance may depend on environmental factors such as metallicity, the spectral shape and intensity of the radiation field, gas density and temperature, and the presence of interstellar shocks. In addition to production in asymptotic giant branch stars, PAH formation and growth in the ISM are thought to be governed by the accretion of gas-phase carbon and the shattering of large carbonaceous grains (S.-J. Rau et al. 2019). Efficient accretion requires sufficiently high metallicities and moderate-to-high gas densities (J. Y. Seok et al. 2014; G. Aniano et al. 2020; C. M. Whitcomb et al. 2024; C. M. Whitcomb et al. 2025; X. Zhang et al. 2025). Moderate shock speeds ($v \lesssim 100$ km s$^{-1}$) may replenish the small PAH population through shattering (potentially providing another channel for small PAH production besides accretion; J. Y. Seok et al. 2014; D. Narayanan et al. 2023). Faster shocks, on the other hand, are sources of PAH destruction (E. R. Micelotta et al. 2010b). PAH destruction is also efficient in the ionized ISM. Photodestruction leads to a dearth of PAHs in HII regions (J. Chastenet et al. 2019, 2023; O. V. Egorov et al. 2023, 2025; J. Chastenet et al. 2025; M. J. Rodríguez et al. 2025). Collisions with atoms and electrons in ionized gas can erode PAHs on timescales of $\tau_{\rm sputter} \sim$ few thousand years for very hot temperatures ($T \gtrsim 10^6$ K). At cooler temperatures ($T \lesssim 10^4$ K) and denser mediums ($n \sim 10^4$ cm$^{-3}$), this timescale can extend to $\tau_{\rm sputter} \sim 10$ Myr (E. R. Micelotta et al. 2010a). Accretion and shattering processes should increase the PAH abundance $q_{\rm PAH}$, while destruction processes (or removal by coagulation onto larger grains) should decrease it (S.-J. Rau et al. 2019).

Given these factors (and that PAH processing has been observed in nearby galaxy halos; J. Chastenet et al. 2026, submitted), how can $q_{\rm PAH}$ remain mostly constant over $\sim$20 Myr in M82 (Figure 9)? Possible explanations include: (1) PAHs are destroyed at the same rate in which they are formed; (2) PAHs are efficiently replenished from the interiors of clouds; and (3) PAHs are undergoing very minimal processing by the hot wind due to environmental shielding. For these explanations to be plausible, certain conditions must be met.

#### 6.2.1. PAHs Are Undergoing Very Little Processing

In the scenario in which PAHs are shielded, the density of the cool clouds must be sufficient to shield the embedded PAHs from photodestruction; however, even moderately dense ($n \sim 100$ cm$^{-3}$) clouds are conducive environments for gas-phase accretion onto PAHs on short timescales ($\sim$1–2 Myr; X. Zhang et al. 2025), which would tip the size distribution toward larger sizes. Therefore, maintaining overall uniform distributions of size and $q_{\rm PAH}$ would require PAHs to be embedded in clouds that are insufficiently dense to promote PAH growth. These PAHs must also not be so embedded that UV photons cannot excite them. For reference, the CO clouds measured in the M82 wind contain number densities ranging $\sim$10–50 cm$^{-3}$ (N. Krieger et al. 2021). Cloud radius is also thought to play a role in dust survival, with larger clouds providing better shielding than smaller clouds since the cloud-crushing timescale grows with increasing radius (R. J. Farber & M. Gronke 2022).

For PAHs to be carried to M82's CGM, their cloud hosts must survive travel times of at least $\sim$20 Myr. Recent JWST observations of the Makani Galaxy provide evidence for PAHs surviving hot galactic outflows out to the CGM ($\sim$35 kpc, or $\sim$100 Myr; S. Veilleux et al. 2025), likely enabled by environmental shielding within cool clouds whose own survival is prolonged by condensation from wind–cloud interactions. Wind-tunnel simulations support this theory (e.g., F. Heitsch & M. E. Putman 2009; E. E. Schneider & B. E. Robertson 2017; M. Gronke & S. P. Oh 2018; M. W. Abruzzo et al. 2022; R. J. Farber & M. Gronke 2022; Z. Chen & S. P. Oh 2024; H. M. Richie et al. 2024), suggesting that radiative cooling faster than the "cloud-crushing" timescale must be invoked in order for cool clouds to survive hot winds. In this scenario, a $T \lesssim 10^4$ K cloud advects material from the hot wind, creating a mixing layer of "warm" gas that rapidly cools to sufficiently replenish the cold cloud (D. B. Fielding & G. L. Bryan 2022). In the case of dusty clouds, smaller clouds lead to faster cloud-crushing timescales and thus prolonged exposure of dust grains to the hot wind, resulting in small-grain destruction (H. M. Richie et al. 2024). This means that the PAH-rich clouds in the M82 wind must be efficiently cooling to survive their journey to the CGM, and there exists cool neutral gas out to these distances (A. K. Leroy et al. 2015; P. Martini et al. 2018; N. Krieger et al. 2021).

If PAH-rich clouds are cooling and mixing with warmer phases of the wind, then PAH emission should be associated with multiple gas phases. Observations support this idea. Out to about ~300 pc from the base of the M82 wind, there is a spatial correlation between 3.3 $\mu$m PAH emission and Pa-$\alpha$ (A. D. Bolatto et al. 2024; D. B. Fisher et al. 2025). This result is corroborated when analyzing the full JWST mosaics (shown in Figure 3), as S. Lopez et al. (2026) found that H$\alpha$ scales with 7.7 and 11.3 $\mu$m PAH emission over a vertical distance of ~2.5 kpc. PAHs are also somewhat correlated with CO-bright gas over this same distance (V. Villanueva et al. 2025a, 2025b; S. Lopez et al. 2026). X-rays, however, are not obviously cospatial with PAH emission. The morphology of prominent arc-like regions in the south can be described as X-ray-filled bubbles surrounded by a leading layer H$\alpha$ and then PAHs (see the Appendix of S. Lopez et al. 2026).

With these considerations, PAHs in the M82 wind are likely embedded in cool clouds whose surfaces are being ionized by UV photons, consistent with PDRs. PAHs are fundamental components of PDRs (e.g., G. C. Sloan et al. 1997; M. Rapacioli et al. 2006; B. Fleming et al. 2010; H. Andrews et al. 2015), contributing to gas heating through the photoelectric effect (E. L. O. Bakes & A. G. G. M. Tielens 1994) and also influencing PDR chemistry by mediating charge exchange with the surrounding gas, thereby neutralizing gas-phase metals (E. L. O. Bakes & A. G. G. M. Tielens 1998). NIRSpec spectroscopy of the Orion Bar and a PDR in the Small Magellanic Cloud have independently determined that PAH intensity peaks in the neutral gas right outside the ionization front (E. Peeters et al. 2024; I. Y. Clark et al. 2025). These results agree with PAHs being embedded in the atomic surface layers of cool clouds.

#### *6.2.2. Efficient PAH Replenishment from Cloud Interiors*

As previously described in this section, a plausible scenario underlying these observations is that PAHs are embedded in the surface layers of cool clouds, allowing them to be illuminated by UV photons from the starburst, giving rise to the observed emission. In this picture, these surface layers would envelop optically thick cloud interiors, likely containing reservoirs of cool gas, dust, and PAHs. Since cloud interiors provide efficient shielding from processing by the supernova wind and ionizing photons, the interior PAH properties likely remain relatively unchanged throughout the outflow region covered by our observations. Therefore, another explanation for the lack of observed PAH processing is that cloud surface layers are constantly being replenished with "fresh" PAHs from cloud interiors.

Theoretical studies suggest that the dominant acceleration mechanism of clouds in outflows is hydrodynamical cloud–wind mixing—primarily Kelvin–Helmholtz instabilities initiated by shear at the cloud–wind interface. The growth of these instabilities to cloud scales occurs in a cloud-crushing time, $t_{\rm cc} = \chi^{1/2} r_{\rm cl}/v_{\rm rel}$, where $\chi$ is the ratio between cloud and wind density, $r_{\rm cl}$ is the initial radius of the (spherical) cloud, and $v_{\rm rel}$ is the relative velocity between the cloud and the wind (R. I. Klein et al. 1994). As these instabilities grow, they dredge up material from cloud interiors, bringing it to cloud surfaces. This process could thereby supply fresh PAHs to the surfaces of clouds, potentially leading to an unchanged appearance of these clouds in PAH emission.

The growth of these instabilities and subsequent replenishment of PAHs may occur on short enough timescales that evolution in PAH properties may not be noticeable within the inner ~2 kpc of the outflow. Clouds in outflows span a wide range in sizes (O. Warren et al. 2025), but for a typical clump size in the inner outflow of M82, $r_{\rm cl} \sim 10$ pc, the cloud-crushing time is roughly 3 Myr (D. B. Fisher et al. 2025). This timescale is at most comparable to, but likely shorter than, the timescale for destruction of PAHs in cloud–wind mixing layers (see Figure 15 of H. M. Richie & E. E. Schneider 2026), meaning that PAHs would be replenished in mixing layers more rapidly than they are destroyed. It is possible that some of the scatter in the profiles may arise from the variation in cloud size, which, for larger clouds, could lengthen $t_{\rm cc}$ beyond the PAH destruction timescale. However, for smaller clouds (which are likely more common in outflows; O. Warren et al. 2025), PAH replenishment would occur so rapidly that PAH evolution trends would likely not be observable on ~kpc scales in the outflow.

Eventually, after enough cloud–wind mixing has occurred, the reservoir of pristine PAHs in cloud interiors would be exhausted, which would lead to observable trends in PAH evolution on the surfaces of clouds. Using our conservative assumption for the cool phase velocity, these observations show at most 7 $t_{\rm cc}$ of cloud evolution, which is a fairly small window into the clouds' overall evolution. Thus, dilution of pristine PAHs in cloud interiors due to cloud–wind mixing, and the resulting evolution in the properties of surface layer PAHs, may not yet be visible in the inner ~2 kpc of the wind probed by our data. Imaging beyond the inner region of the outflow would reveal whether PAH evolution takes effect on larger scales, and would therefore be helpful in clarifying the likelihood of this physical scenario.

#### *6.2.3. An Equilibrium between PAH Destruction and Formation Processes?*

A balance between PAH destruction and formation mechanisms is unlikely for several key reasons. To start, the destruction and formation of PAHs are both highly dependent on environment. For example, in protoplanetary disks, softer UV photons can only dissociate PAH molecules within $r < 1$ au of the host star, whereas EUV and X-ray photons will dissociate PAHs throughout the entire disk (R. Siebenmorgen & E. Krügel 2010). This is observational evidence for PAH destruction being reliant on the radiation field hardness, which is also well-documented in galaxies and AGNs (e.g., T. Monfredini et al. 2019; D. Rigopoulou et al. 2021; T. S.-Y. Lai et al. 2023; O. V. Egorov et al. 2025) and supported by theory (e.g., B. T. Draine et al. 2021; D. Rigopoulou et al. 2021; H. M. Richie & B. S. Hensley 2025). As discussed in the previous subsection, the spectral shape and intensity of the radiation field may change with distance, subsequently changing the rate of PAH destruction. To keep $q_{\rm PAH}$ constant, any change in the destruction rate must be counteracted with a change in the formation rate. Because PAH formation and growth is also environmentally dependent (e.g., requiring moderate-to-high gas densities and solar metallicities), this immediate balance seems implausible in the wind of M82.

Additionally, a balance between PAH destruction and formation would require these processes to occur on similar timescales. Using the chemical evolution models presented in R. S. Asano et al. (2013), J. Y. Seok et al. (2014) investigated shattering (formation) and coagulation (destruction) timescales, which are both relevant at solar metallicity, in various ISM phases. They found that both timescales are dependent on the dust-to-gas ratio, and that coagulation is only efficient in dense molecular clouds ($T \sim 10$ K), a regime in which shattering is inefficient. Conversely, $\tau_{\rm shatter}$ outpaces $\tau_{\rm coag}$ at temperatures $T \sim 10^4$ K, consistent with the temperatures of the gas phases probed in this study. These timescales hold true for both J. S. Mathis et al. (1977) and J. C. Weingartner & B. T. Draine (2001b) grain size distributions. Given these points, it appears very difficult to finely tune PAH formation and destruction mechanisms to balance each other out and create a steady $q_{\rm PAH}$ over time.

#### *6.2.4. Abundance of PAH-sized Grains in a Simulated M82-like Outflow*

Recently, H. M. Richie & E. E. Schneider (2026) produced high-resolution hydrodynamic simulations of dusty galactic winds. Achieving both similar spatial resolution and coverage as the JWST data of this work, these simulations present the unique opportunity to directly compare our observations with a simulated M82-like wind on comparable physical scales and timescales. In detail, H. M. Richie & E. E. Schneider (2026) leverages the Cholla hydrodynamics code (E. E. Schneider & B. E. Robertson 2015) to model a $z \sim 0$ galaxy with a nuclear starburst concentrated in a 300 pc radius with a star formation rate of 5 $M_\odot$ yr$^{-1}$. These conditions replicate the environment found in M82. The simulations also incorporated M82-specific stellar, gas, and dark matter masses, as well as the cluster mass function observed in M82 (N. McCrady & J. R. Graham 2007; Y. D. Mayya et al. 2008; J. P. Greco et al. 2012; R. C. Levy et al. 2024). The simulations are run over a timescale of 40 Myr. This is enough time for stellar feedback to drive a hot wind that sweeps dusty cool material out of the disk, accelerating it $\sim$10 kpc toward the halo. In Figure 10 we plot snapshots 15 Myr after the initial burst, when cool material has only accelerated to $z \sim 3$ kpc. Both panels represent the surface density of PAH-sized ($r = 0.001\ \mu$m) and larger ($r = 0.1\ \mu$m) grains, assuming spherical grains.

With these simulations, we perform a similar PAH abundance experiment as in Section 5. Here, we model the abundance of PAH-sized dust grains as the ratio of the surface density of 0.001 $\mu$m grains to the surface density of 0.1 $\mu$m grains (the latter of which is where the large grain population roughly peaks). An important caveat is that while the 0.001 $\mu$m grains are PAH-sized ($\sim$500 carbon atoms), they are not actually modeled as PAHs, and true PAH destruction may be more efficient than the sputtering of spherical dust grains (E. R. Micelotta et al. 2010a). Therefore, the destruction rates of PAH-sized grains in H. M. Richie & E. E. Schneider (2026) serve as upper limits. Thus, we can think of the abundance of 0.001 $\mu$m grains with respect to 0.1 $\mu$m grains as an upper limit for $q_{\rm PAH}$, which we will hereafter refer to as $q_{\rm PAH}^{\rm UL}$. Note that in this section we refer to 0.001 $\mu$m grains as PAHs, although they are technically modeled as spherical dust grains.

Following Section 5, we average the distribution of 0.001 and 0.1 $\mu$m grains down a 33″-wide slice (white rectangles in the first two panels of Figure 10). We then plot $q_{\rm PAH}^{\rm UL}$ as a function of vertical distance from the midplane. Because the simulations assume an arbitrary fraction of PAHs at $z = 0$ kpc and $t = 10$ Myr, we normalize $q_{\rm PAH}^{\rm UL}$ by this baseline value. We have the advantage of evolving $q_{\rm PAH}^{\rm UL}$ with time (color bar in the third panel). The central $\pm$1 kpc is masked out, as in Figure 8. It is important to note that Figure 10 shows projections of the simulated dust densities. If the PAH emission in our observed data does indeed arise from the surfaces of clouds, then the simulated $q_{\rm PAH}$ profiles show an importantly different quantity than the observed profiles. Specifically, the $q_{\rm PAH}^{\rm UL}$ profiles show the *total* PAH abundance in the projected *xz*-plane, while the observed profiles may only probe $q_{\rm PAH}$ of the cloud surfaces. In this case, the two profiles would likely exhibit a differing $z$–dependence, even with the same underlying PAH properties, due to geometric effects. This is because $q_{\rm PAH}^{\rm UL}$ only probes evolution in the PAH abundance, while $q_{\rm PAH}$ (a function of cloud surface area) also probes cloud size. Because clouds increase in volume as they expand radially outward into the outflow, their surface areas increase with $z$, which could lead to higher PAH fluxes. These considerations limit absolute direct comparison with the observed profiles. However, the simulations still provide a useful framework to assess how the underlying PAH population may be evolving in the M82 outflow.

Initially, $q_{\rm PAH}^{\rm UL}$ shows a strong decline with vertical distance. In the first 10 Myr, the 0.001 $\mu$m grains are still concentrated close to the disk and have not had enough time to travel to farther distances. This is because 0.001 $\mu$m grains are rapidly destroyed in $T \gtrsim 10^4$ K gas and, thus, are strongly confined to cool clouds in the outflow. As such, the buildup of these grains at large distances takes considerable time, since the acceleration of initially stationary cool gas from the disk occurs gradually (via ram pressure acceleration and hydrodynamical mixing with the supernova wind). By the 30 Myr mark, the wind has built up enough cool gas and PAH-sized grains to populate the entire simulation volume. Another driver of the relatively steep slope in $q_{\rm PAH}^{\rm UL}$ at early times is the lack of environmental shielding for dusty cloud material, since the disk is initially surrounded by a hot virialized halo containing no cool gas. As dusty clouds expand into the hot halo from the disk, dusty gas that is stripped from clouds by cloud–wind mixing has relatively few opportunities to recondense onto cool clouds, leading to overall higher rates of dust destruction.

Over the course of 10–22.5 Myr, the $q_{\rm PAH}^{\rm UL}$ profiles exhibit a simultaneous decrease in the normalization and a flattening with distance. We interpret these behaviors as follows. The decreasing normalization of $q_{\rm PAH}^{\rm UL}$ is the result of a declining PAH-to-dust mass ratio, likely due to PAH destruction by supernova feedback in the disk.[35] The $q_{\rm PAH}^{\rm UL}$ profiles becoming flatter may be due to two effects. First, as described earlier in this section, enough time has passed for cool gas (which shields the 0.001 $\mu$m grains) moving at a speed of roughly 200 km s$^{-1}$ to expand out to a distance of 5 kpc. Second, the fraction of cool gas in the outflow increases at later times because of the existence of material from earlier ejections, providing more efficient environmental shielding for the dust.

These effects have interesting implications on the observations of M82 presented in this paper. In particular, if the declining normalization of the $q_{\rm PAH}^{\rm UL}$ profiles are due to supernova destruction, then the 0.001 $\mu$m grains must not be replenished on short timescales in the disk. Therefore, supernova feedback in the starburst can drive down the amount of available PAH-sized grains that can eventually be launched in the wind. This aligns well with the low $q_{\rm PAH}$ ($\sim$1%) that we measured in Section 5. The flat $q_{\rm PAH}$ profiles observed in M82 could also indicate that the current burst period has depleted some PAHs in the star-forming region of M82's disk. The flattening of the profiles through the buildup of PAHs at high vertical distance indicates that PAHs have had enough time to populate the outflow volume. For PAHs to have made it to the halo, environmental shielding by entrained cool clouds must have occurred, as discussed earlier in this section. Additionally, finding PAHs on vertical scales of 5 kpc can only be explained by (1) a continuous burst of $\sim$25 Myr that drives material at a speed of $\sim$200 km s$^{-1}$, or (2) multiple starburst episodes. Taking into consideration M82's star formation history, the latter seems the most plausible, as M82 is thought to have had two distinct bursts about $\sim$10 and $\sim$5 Myr ago (N. M. Förster Schreiber et al. 2003). With this in mind, the current wind is expanding into a dusty halo that has already been enriched with PAHs from previous bursts. Therefore, the observed PAH abundance in the inner M82 wind bears the imprint of multiple starburst episodes.

[35] H. M. Richie & E. E. Schneider (2026) noted a simulation artifact where densities are high enough ($n \sim 200$ cm$^{-3}$) in the disk that sputtering is artificially enhanced, which decreases the initial dust-to-gas ratio and therefore limits the available dust that could be entrained in the outflow. However, this artifact plays an insignificant role in the 0.001 $\mu$m dust grains and thus should not impact these results.

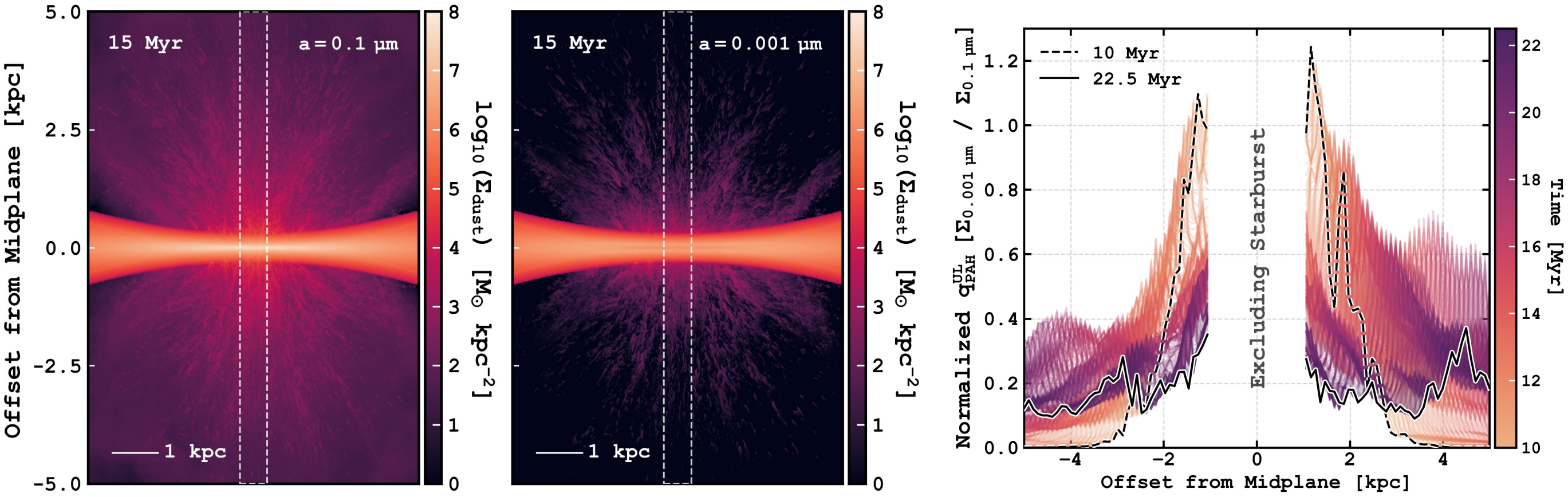

**Figure 10.** The evolution of the abundance of PAH-sized grains computed from the hydrodynamical simulations of H. M. Richie & E. E. Schneider (2026). This abundance is approximated as the fraction of the surface density of 0.001 $\mu$m (PAH-sized) to 0.1 $\mu$m grains, assuming spherical dust grains. This calculation can be considered an upper limit on PAH abundance, hereafter $q_{\rm PAH}^{\rm UL}$. The left and middle panels show the distribution of these grains in an M82-like dusty outflow 15 Myr after the initial burst. We average the surface density of the dust grains down a 33″-wide slice (white rectangles) whose area matches the slices of Figure 8. The right panel shows $q_{\rm PAH}^{\rm UL}$ as a function of offset from the disk midplane, normalized to the initial value at $z = 0$ kpc and $t = 10$ Myr. Over the course of 10–22.5 Myr (color bar), the $q_{\rm PAH}^{\rm UL}$ profiles flatten while their normalization decreases. The decreasing normalization suggests that, in the disk, supernova feedback limits the amount of PAHs available for launch in the outflow. Meanwhile, the flattening of the profiles may be due to two simultaneous effects: (1) PAH-sized grains entrained in the wind are being shielded by cool clouds, and (2) already-existing PAH-sized grains are populating the halo from previous bursts (see discussion in Section 6.2). These results are consistent at both PACS 70 and 160 $\mu$m resolutions.

Caveats to directly comparing these simulations with our observations are as follows. First, we re-emphasize that the spherical 0.001 $\mu$m dust grains modeled in H. M. Richie & E. E. Schneider (2026) are not true PAHs, and thus, $q_{\rm PAH}^{\rm UL}$ is an upper limit to an actual measurement of $q_{\rm PAH}$. Similarly, these simulations do not yet consider shattering, which may replenish the small PAH population. Finally, the timescales we are quoting in our observations are dependent on an assumed constant velocity of $v = 200$ km s$^{-1}$, informed by typical speeds of cold galactic wind phases (e.g., A. D. Bolatto et al. 2013). The H. M. Richie & E. E. Schneider (2026) simulations more accurately accelerate the outflow rate with distance, with the average speed of a few hundred kilometers per second. As a result, there is a slight mismatch between the mapping of evolutionary time and distance between our work and H. M. Richie & E. E. Schneider (2026).

## 7. Summary and Conclusions

In this paper, we present new JWST images of the cool phase of the M82 galactic superwind, as traced by PAHs. These observations map the inner $\sim$5 kpc, including the starbursting disk, at $\sim$0″.05–0″.375 ($\sim$1–6.5 pc) resolution. The combination of JWST's sensitivity, angular resolution, and unique access to the 3.3 $\mu$m PAH complex provides powerful new diagnostics of cool material in galactic winds. With these observations, our goal is to track the state of this material as it is entrained by the hot wind, lifted from the disk, and transported toward the CGM. We summarize our findings as follows.

1. PAH surface brightness scales with the inverse square function of vertical distance ($z^{-2}$), suggesting that variation in PAH intensity is largely due to distance from the incident radiation field of the starburst (Figures 3 and 4).
2. Within a factor of $\sim$2, the 3.3/11.3 and 3.3/7.7 $\mu$m ratios are mostly flat throughout the wind, while the 11.3/7.7 $\mu$m ratio shows a modest increase with distance outside the starburst. The broad silicate absorption feature at 9.7 $\mu$m underestimates the intensity of the 11.3 $\mu$m complex in the disk, which biases the 11.3/7.7 $\mu$m (3.3/11.3 $\mu$m) ratio toward lower (higher) values. (Figures 5 and 6).
3. Comparison with the dust models from B. T. Draine et al. (2021) reveals that PAHs range between standard-to-high ionization states and standard-to-large sizes. PAHs are becoming somewhat less ionized (and possibly larger) with distance from the starburst, with charge evolution being the more significant trend, as PAHs are exposed to a declining radiation field and ionization parameter with distance. These trends are still evident, though with reduced significance, when measured over a larger azimuthal area and thus a more global scale. The strength and hardness of the radiation field may be the dominant contributor to the observed band ratios (Figure 7).

4. In the wind, PAH abundance is low at $q_{\mathrm{PAH}} \sim 1\%$, likely set by the intense radiation field and shocks in the starburst. Within a factor of ∼2, PAHs that are successfully launched remain largely unchanged over ∼20 Myr (assuming outflow speeds of 200 km $s^{-1}$) out to $d \sim 5$ kpc ($\theta \approx 4.8'$), suggesting that they are protected by being embedded in the surfaces of cool clouds. Rapid mixing with the hot wind and possible replenishment from cloud interiors may allow PAH-rich material to survive out to the halo (Figures 8 and 9).
5. Simulations of an M82-like galactic outflow reveal that a flat $q_{\mathrm{PAH}}$ may also be the consequence of the wind expanding into a halo populated with dust from previous bursts. The low normalization of $q_{\mathrm{PAH}}$ is likely set by PAH destruction from supernova feedback in the disk (Figure 10).

This work only scratches the surface of what can be learned about the M82 wind in the era of JWST. New studies are in the works to diagnose the relationship between the observed 3.3 $\mu$m PAH emission and the radiation field (P. Arens et al. 2026, in preparation) and to quantify the filamentary structure of the wind (S. Cronin et al. 2026, in preparation). Wider-field JWST imaging of M82 (i.e., capturing the total extent of PAH emission in Figure 1) is needed to trace the fate of this cool material at parsec-scale resolution, as it is carried into the halo, and NIRSpec spectroscopy could be the key to unlocking the true timescales in which this cool material survives the wind. Future observations are essential to inform theory and simulations that are attempting to replicate the detailed structure and physics of dusty galactic winds.

## Acknowledgments

We thank the anonymous referee for providing a helpful report and Ryan Chown for insightful discussions that improved this paper. We gratefully acknowledge Alyssa Pagan (STScI) and ESA/Webb, NASA, and CSA for producing the beautiful images presented in Figure 2. This research benefited from engaging discussions at the 5th Pan-Dust Conference, held at the University of Arizona in 2025 November.

This work is based on observations made with the NASA/ESA/CSA JWST. The data were obtained from the Mikulski Archive for Space Telescopes at the Space Telescope Science Institute, which is operated by the Association of Universities for Research in Astronomy, Inc., under NASA contract NAS 5-03127 for JWST. All JWST data used in this paper can be found in MAST: DOI: 10.17909/repd-an86. These observations are associated with program JWST-GO-01701. Support for program JWST-GO-01701 is provided by NASA through a grant from the Space Telescope Science Institute, which is operated by the Association of Universities for Research in Astronomy, Inc., under NASA contract NAS 5-03127. This research has made use of NASA's Astrophysics Data System Bibliographic Services. This research used resources of the Oak Ridge Leadership Computing Facility, which is a DOE Office of Science User Facility supported under contract DE-AC05-00OR22725, using Frontier allocation AST181 and SummitPLUS allocation AST200.

S.A.C., A.D.B., and Y.-H.T. acknowledge support from the NSF under award AST-2108140. V.V. acknowledges support from the Comité ESO Mixto 2024 and from the ANID BASAL project FB210003. I.D.L. acknowledges funding from the Belgian Science Policy Office (BELSPO) through the PRODEX project "JWST/MIRI Science Exploitation" (C4000142239), and funding from the European Research Council (ERC) under the European Union's Horizon 2020 research and innovation program DustOrigin (ERC-2019-StG-851622). This research was carried out in part at the Jet Propulsion Laboratory, California Institute of Technology, under a contract with the National Aeronautics and Space Administration (80NM0018D0004).

*Facilities:* JWST (MIRI, NIRCam), Spitzer (IRAC, IRS), Herschel (PACS, SPIRE).

*Software:* `astropy` (Astropy Collaboration et al. 2013, 2018), `numpy` (S. Van Der Walt et al. 2011; C. R. Harris et al. 2020), `matplotlib` (J. D. Hunter 2007), `scipy` (P. Virtanen et al. 2020), `jupyter` (T. Kluyver et al. 2016), JWST Calibration Pipeline (H. Bushouse et al. 2023, 2024), `STPSF` (M. Perrin et al. 2024), `DOLPHOT` (A. E. Dolphin 2000; A. Dolphin 2016; D. R. Weisz et al. 2024), `pypher` (A. Boucaud et al. 2016), `PAHFIT` (J. D. T. Smith et al. 2007), `JHAT` (A. Rest et al. 2023), `synphot` (STScI Development Team 2018).

## Appendix A
## Continuum Maps

We present the continuum maps referenced in Sections 2.6 and 2.7 in Figure 11. The leftover continuum after subtracting the 3.3 $\mu$m PAH component from the F335M image is presented in the left panel. We determine a clean continuum removal as one that does not leave obvious artifacts (e.g., over-subtracted pixels in the 3.3 $\mu$m map; Figure 3) nor obvious outflow emission in the continuum map. The resulting continuum map is that of a galaxy disk dominated by starlight, with possible extinction due to dust lanes. The middle and right panels of Figure 11 show the result of `PAHFIT` modeling of available Spitzer IRS data (see Section 2.7 for details). With these maps, we perform a pixel-by-pixel subtraction of the continuum (dominated mostly by thermal dust continuum and some starlight) underlying the F770W and F1130W images. For the ∼25% of the MIRI maps that do not have corresponding Spitzer IRS coverage, we subtract median percentages determined in the Spitzer maps: ∼14% (wind) and ∼16% (disk) for F770W, and ∼18% (wind) and ∼22% (disk) for F1130W.

Some artifacts exist in the 7.7 $\mu$m continuum map, including ring-like structures and a few blank pixels in the disk where `PAHFIT` modeling failed. We fill in the bad pixels with the mean of the nearest neighbors. We verify that this interpolation and the "ringing" do not result in artifacts in the 7.7 $\mu$m PAH map (Figure 3).

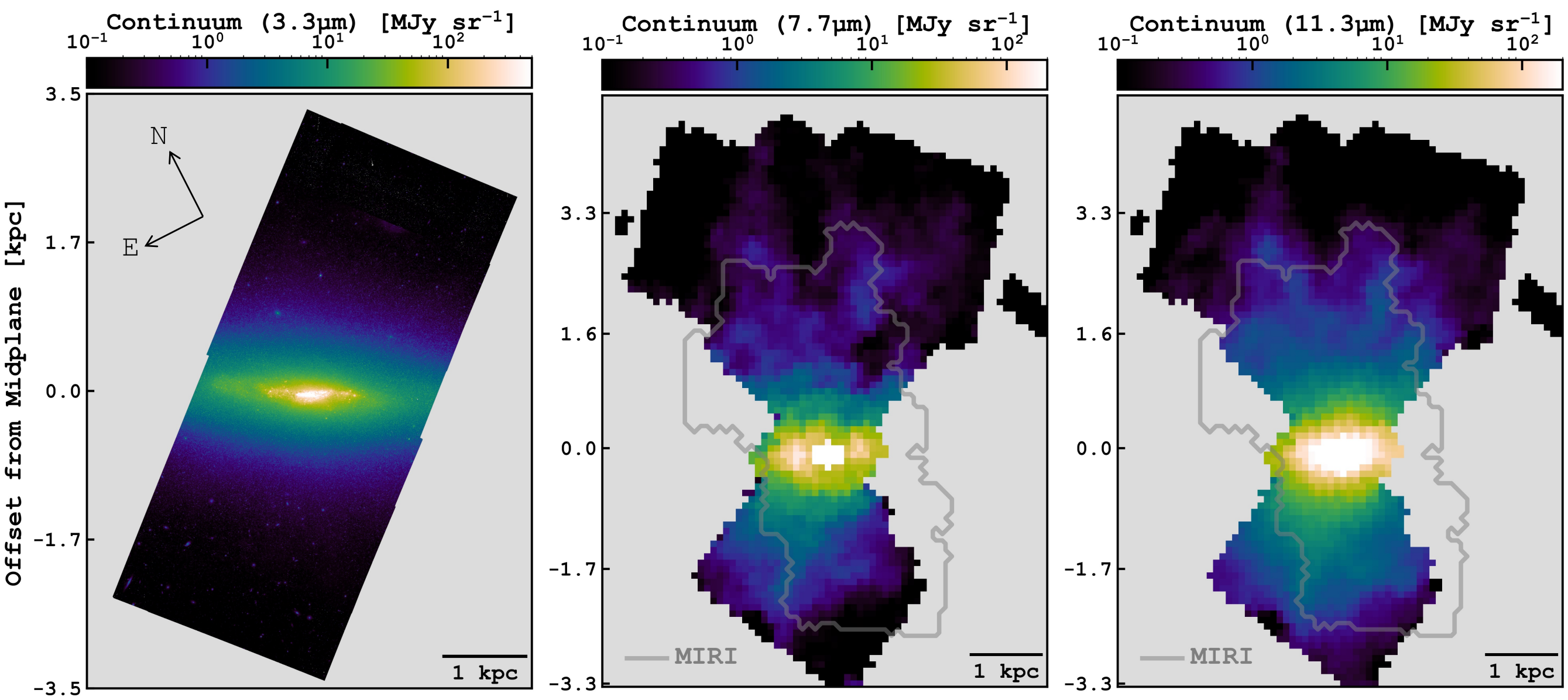

**Figure 11.** Continuum estimates subtracted from the F335M, F770W, and F1130W images to produce the 3.3, 7.7, and 11.3 $\mu$m maps in Figure 3. The continuum underlying the 3.3 $\mu$m feature (dominated by starlight) is isolated using an iterative approach presented in Section 2.6. The left panel is the result of this continuum estimation at the resolution of the F360M image ($0\farcs12 \approx 2$ pc). The middle and right panels show the continuum estimates derived by PAHFIT modeling of available Spitzer IRS data (Section 2.7). The gray contours outline the MIRI footprint, degraded to the $\approx 5''$ pixel scale of the Spitzer maps.

## Appendix B
## 11.3/7.7 $\mu$m PAH Ratio with Spitzer

We present the full 11.3/7.7 $\mu$m ratio map from Spitzer IRS spectroscopy in Figure 12. The intensities ($F_\nu$ [MJy sr$^{-1}$]) used in this ratio were derived by applying synthetic photometry to mosaicked Spitzer IRS cubes (see Section 2.7 for a description of the synthetic photometry). Before taking the ratio, we subtract the continuum from the synthetic photometry using the maps in Figure 11. Ring-like features arise around the starburst from extended wings caused by the PSF on such a bright source. After projecting the 11.3/7.7 $\mu$m ratio onto the MIRI grid, we compare the Spitzer radial profiles of the 11.3/7.7 $\mu$m ratio (averaged down 33$''$-wide rectangle in Figure 12) with the profiles measured from JWST imaging in Figure 6. The "ringing" features manifest as oscillatory fluctuations, but overall, the general trend is comparable to the results derived with JWST photometry: the 11.3/7.7 $\mu$m ratio dips toward the disk due to silicate absorption affecting the 11.3 $\mu$m complex, and more or less evens out in the wind.

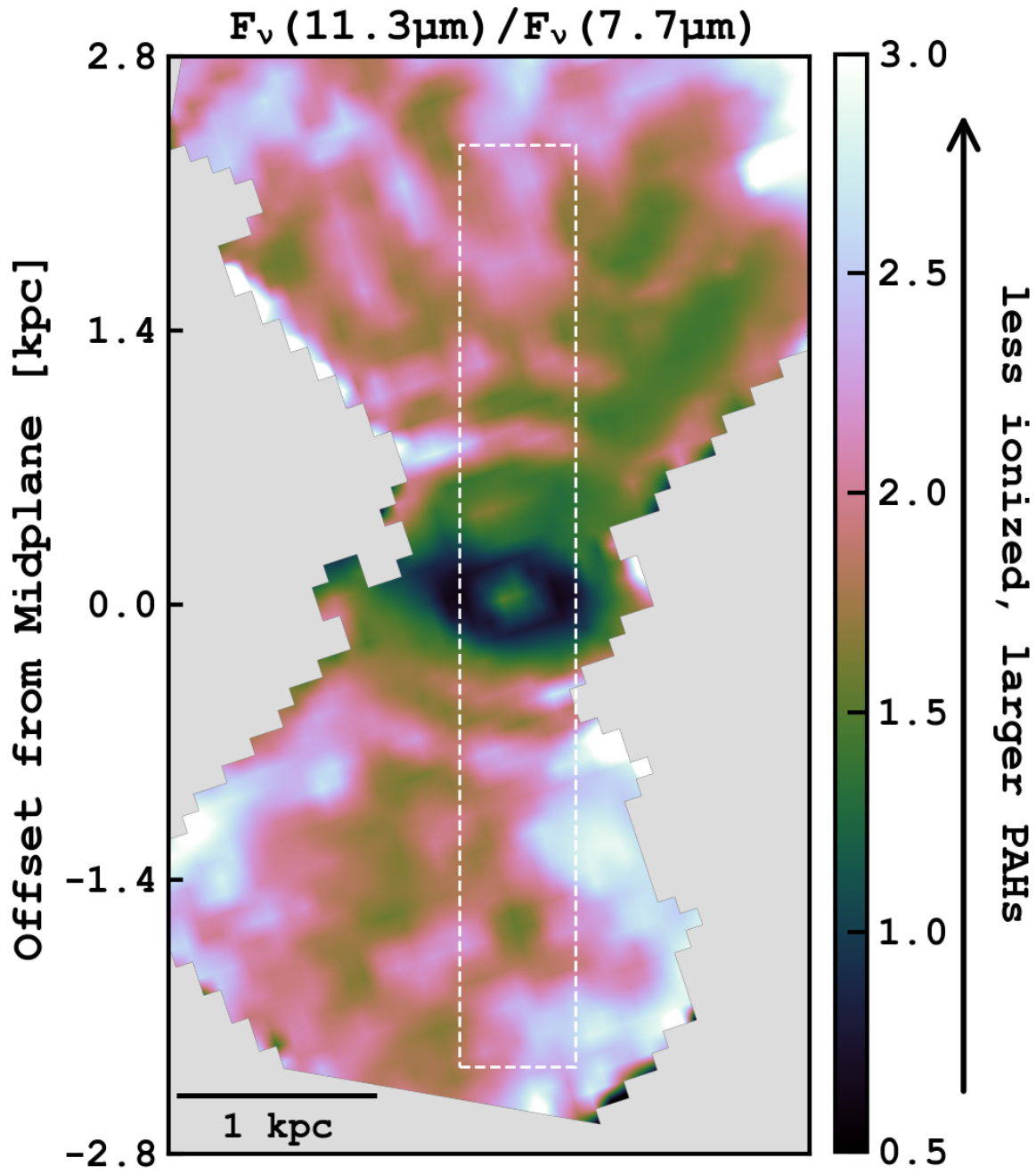

**Figure 12.** The 11.3/7.7 $\mu$m ratio throughout the M82 wind estimated from synthetic photometry applied to Spitzer IRS data. We project the data onto the MIRI coordinate grid. The "ringing" features are PSF-driven artifacts caused by extended wings associated with the bright starburst. These artifacts manifest as oscillations in the radial profile of the ratio (see Figure 6). Overall, the general trend matches what is measured with JWST photometry in this work.

## Appendix C
## Corrections Applied to Spitzer IRAC Channel 4 Image

To properly compare the luminosities of the 7.7 $\mu$m PAH complex as measured by Spitzer IRAC4 and F770W (Section 5), corrections to the Spitzer data are required. We first apply the extended-source correction for an infinite aperture at 8 $\mu$m (0.74; see Table 8.2 in the IRAC Instrument Handbook). To estimate the color correction between IRAC4 and F770W, we apply synthetic photometry to the PDRs4All templates of the Orion Bar (R. Chown et al. 2024) using the IRAC4 and F770W spectral response curves and synphot. We calculate the F770W/IRAC4 color correction for each Orion Bar region to be: 0.94 (H II), 0.99 (APDR), and 1.05 for DF1, DF2, and DF3. Because Figure 7 shows that the physical state of the PAHs in the M82 wind is closest to the H II region of the Orion Bar, we choose to apply the H II region color correction. Finally, we convert the Spitzer photometry to match the JWST photometric convention of assuming a flat $F_\nu$ spectrum; Spitzer instead assumes an underlying flux density $F_\nu \propto \nu^{-1}$. These corrections produce a flux density measured in IRAC4 that matches within $\sim$10% of F770W.

## ORCID iDs

Serena A. Cronin https://orcid.org/0000-0002-9511-1330
Alberto D. Bolatto https://orcid.org/0000-0002-5480-5686
Helena M. Richie https://orcid.org/0000-0001-6325-9317
Grant P. Donnelly https://orcid.org/0009-0001-6065-0414
Rebecca C. Levy https://orcid.org/0000-0003-2508-2586
Karl D. Gordon https://orcid.org/0000-0001-5340-6774
Elizabeth Tarantino https://orcid.org/0000-0003-1356-1096
Martha L. Boyer https://orcid.org/0000-0003-4850-9589
Lee Armus https://orcid.org/0000-0003-3498-2973
Patricia A. Arens https://orcid.org/0009-0001-6844-9758
Leindert A. Boogaard https://orcid.org/0000-0002-3952-8588
Daniel A. Dale https://orcid.org/0000-0002-5782-9093
Keaton Donaghue https://orcid.org/0009-0004-5807-9142
Bruce T. Draine https://orcid.org/0000-0002-0846-936X
Sara E. Duval https://orcid.org/0000-0003-0014-0508
Kimberly Emig https://orcid.org/0000-0001-6527-6954
Deanne B. Fisher https://orcid.org/0000-0003-0645-5260
Simon C. O. Glover https://orcid.org/0000-0001-6708-1317
Brandon S. Hensley https://orcid.org/0000-0001-7449-4638
Rodrigo Herrera-Camus https://orcid.org/0000-0002-2775-0595
Ralf S. Klessen https://orcid.org/0000-0002-0560-3172
Thomas S.-Y. Lai https://orcid.org/0000-0001-8490-6632
Laura Lenkić https://orcid.org/0000-0003-4023-8657
Adam K. Leroy https://orcid.org/0000-0002-2545-1700
Ashley E. Lieber https://orcid.org/0000-0002-1254-4174
Ilse De Looze https://orcid.org/0000-0001-9419-6355
Sebastian Lopez https://orcid.org/0000-0002-2644-0077
David S. Meier https://orcid.org/0000-0001-9436-9471
Elisabeth A.C. Mills https://orcid.org/0000-0001-8782-1992
Karin M. Sandstrom https://orcid.org/0000-0002-4378-8534
Evan Schneider https://orcid.org/0000-0001-9735-7484
Kaitlyn E. Sheriff https://orcid.org/0009-0005-8382-0614
Utsav Siwakoti https://orcid.org/0000-0003-3633-0098
Evan D. Skillman https://orcid.org/0000-0003-0605-8732
J.D.T. Smith https://orcid.org/0000-0003-1545-5078
Yu-Hsuan Teng https://orcid.org/0000-0003-4209-1599
Todd A. Thompson https://orcid.org/0000-0003-2377-9574
Alexander G.G.M. Tielens https://orcid.org/0000-0003-0306-0028
Sylvain Veilleux https://orcid.org/0000-0002-3158-6820
Vicente Villanueva https://orcid.org/0000-0002-5877-379X
Fabian Walter https://orcid.org/0000-0003-4793-7880
Paul P. van der Werf https://orcid.org/0000-0001-5434-5942